\documentclass[journal]{IEEEtran}

\newfont{\bb}{msbm10 scaled 1100}
\newcommand{\CC}{\mbox{\bb C}}
\newcommand{\RR}{\mbox{\bb R}}
\usepackage{graphicx}
\usepackage{psfrag}
\usepackage{subfigure}
\usepackage{algorithm,algorithmic}
\usepackage{url}
\usepackage{color}
\usepackage{amsmath}
\usepackage{amsfonts,amssymb,amsmath}
\usepackage{mathrsfs}
\usepackage{array}

\begin{document}
\title{{A Tutorial on the Optimization of Amplify-and-Forward MIMO Relay Systems}}
\author{Luca~Sanguinetti, \emph{Member, IEEE}, Antonio A.~D'Amico and~Yue Rong, \emph{Senior Member, IEEE}.\thanks{
\newline \indent {This research was supported in part by the Seamless Aeronautical Networking through
integration of Data links, Radios, and Antennas (SANDRA) project co-funded by the European Commission within the ``Cooperation Programme'' GA No. FP7- 233679.}
\newline \indent Luca Sanguinetti and Antonio A. D'Amico are with the
Department of Information Engineering, University of Pisa, Via
Caruso 56126 Pisa, Italy (e-mail: \{luca.sanguinetti, a.damico\}@iet.unipi.it).
\newline \indent Yue Rong is with the Department of Electrical and Computer Engineering, Curtin University of Technology, Bentley, WA 6102, Australia (e-mail: y.rong@curtin.edu.au).
}}
%

\maketitle

\begin{abstract}
The remarkable promise of multiple-input multiple-output (MIMO) wireless channels has motivated
an intense research activity to characterize the theoretical and
practical issues associated with the design of transmit (source) and receive (destination) processing matrices under different operating conditions.
This activity was primarily focused on point-to-point (single-hop) communications but more
recently there has been an extensive work on two-hop or multi-hop settings in which single or multiple relays are used to deliver the information from the source to the destination. The aim of this tutorial
is to provide an up-to-date overview of the fundamental results and
practical implementation issues of designing amplify-and-forward MIMO relay systems.
\end{abstract}
\begin{keywords}
Tutorial, MIMO,  optimization, transceiver design, amplify-and-forward, non-regenerative relay, power allocation, majorization theory, quality-of-service requirements, single-hop, two-hop, multi-hop, one-way, two-way, multiple relays, perfect channel state information, robust design.
\end{keywords}

\section{Introduction}

\PARstart{M}{ultiple}-input multiple-output (MIMO) relay communications are viewed as one of the most promising techniques to improve the reliability and coverage of wireless systems. 
While the optimization of point-to-point (single-hop) MIMO systems has been widely analyzed (an excellent survey on this topic can be found in \cite{PalomarBook}), the optimization of MIMO relay networks has gained much attention only recently. The aim of this tutorial is to provide an overview of the results obtained in this area. Due to the considerable amount of work in this field and the rapidly intensifying efforts at the time of writing, our exposition will be necessarily incomplete and will reflect the subjective tastes and interests of the authors. To compensate for this partiality, a list of references is provided as an entree into the
extensive literature available on the subject.


{As is well-known, a first operating distinction in relay communications is made on the way the received signals are processed by the relays. This can be done according to several different protocols such as decode-and-forward, amplify-and-forward, compressed-and-forward, mixed-forward and so forth (see for example \cite{Gupta2005} and references therein). The simplest one is the \emph{amplify-and-forward} (AF) protocol in which \emph{non-regenerative} relays are used to linearly process the received signals and to {re-transmit} them toward the destination. Though inherently affected by noise propagation effects, the AF protocol is nowadays considered as the most promising solution for future and/or existing wireless communications since it provides a reasonable trade-off between benefits and practical implementation costs. 
Among the different AF relay systems, the simple two-hop model (in which the information is passed from the source to the destination using one or more \textit{parallel} relays) has been the focus of much ongoing research. For this reason, this tutorial is largely dedicated to the analysis of two-hop architectures while the multi-hop case is reviewed only briefly. In addition, to simplify the exposition, the single relay scenario is almost exclusively considered.} 

{A second operating distinction can be made between \emph{full-duplex} and \emph{half-duplex} systems depending on whether relays can transmit and receive simultaneously or not. This work is focused on half-duplex systems since several practical constraints such as power consumption, implementation costs and spatial efficiency\footnote{Full-duplex systems require an opportune spatial separation between transmit and receive antennas in order to reduce loop-back interference.} make them more appealing for wireless applications.

}

This tutorial is organized as follows\footnote{The following notation is used throughout the paper. Boldface upper and lower-case letters denote
matrices and vectors, respectively, while lower-case letters denote
scalars. We use $\mathbf{A}=\mathrm{diag}\{ a_n\,;\,\,n = 1,2,
\ldots ,K\}$ to indicate a $K \times K$ diagonal matrix with
entries $a_n$ while $\mathbf{A}=\mathrm{diag}\{ \mathbf{A}_1,\mathbf{A}_2,\ldots,\mathbf{A}_K\}$ stands for a block diagonal matrix. The notations ${\bf{A}}^{ - 1}$ and
${\bf{A}}^{ 1/2}$ denote the inverse and square-root of a
matrix ${\bf{A}}$. We use ${\bf{I}}_K$ to denote the identity matrix of order $K$ while $\left[
\cdot \right]_{k,\ell}$ indicates the ($k ,\ell$)th entry of the
enclosed matrix. In addition, we use  $ {\rm{E}}\left\{ \cdot \right\}$ for
expectation, the superscript $ ^T$ and $ ^H$ respectively for transposition and Hermitian transposition.
}. The optimization of a one-way two-hop MIMO system is considered in Section II -- the largest one of this work in view of the considerable attention devoted to such systems. The signal model for linear architectures is first introduced under the assumptions of a frequency-flat propagation channel and a negligible source-destination link. Two different optimization problems are considered. {In particular, the first is focused on the minimization/maximization of a global objective function subject to average power constraints at the source and relay nodes while the second aims at minimizing the total power consumption while satisfying specific quality-of-service requirements}. Next, our analysis is extended to non-linear architectures as well as to frequency selective channels. We also examine the problem of acquiring channel state information at all nodes and describe the distinctive features of some robust optimization solutions. Section III is devoted to the optimization of a multi-hop relay network, while in Section IV a one-way two-hop MIMO system with multiple parallel relays is considered. The direct link is investigated in Section \ref{DirectLink} while the most recently advanced  solutions for the optimization of a two-way two-hop MIMO system are reviewed in Section \ref{Two-wayTwo-hop}. Finally, in Section VII we summarize some interesting open issues that are likely to be the basis for future research in the optimization of relay networks.

\begin{figure}[tbp]
\centering \includegraphics[width=.47\textwidth]{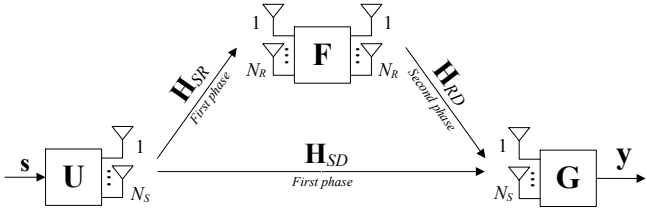}
\caption{Block diagram of a linear one-way two-hop linear MIMO system.}\label{fig1}
\end{figure}

\section{Optimization of a one-way two-hop MIMO system}

The one-way class refers to conventional two-hop systems in which four phases are needed to exchange information between source and destination via the relay link. A first phase is required to convey data from source to relay, a second one is needed to go from relay to destination, and two other phases are then required for reverse link. The block diagram of a one-way two-hop MIMO system in which linear processing is employed at all nodes is shown in Fig. \ref{fig1}. We start considering the case in which the direct link is negligible due to large path attenuation\footnote{This choice is motivated by the fact that a relay plays a much more important role when the direct link is weak than when it is strong.}.
This assumption will be removed in Section \ref{DirectLink}. For simplicity, we also restrict our attention to a frequency flat-fading channel while the extension to frequency-selective fading channels will be discussed only later.

\subsection{Signal model}

The $k$th symbol is denoted by $s_k$ and is taken from a quadrature amplitude modulation (QAM) constellation with an average power normalized to unity for convenience. We assume that the source and destination are both equipped with $N_S$ antennas while the relay employs $N_R$ antennas\footnote{The results can be easily extended to a more general case in which different number of antennas are available at source and destination.}. In addition, we denote by $K \le \min{(N_R, N_S)}$ the total number of transmitted symbols.

As shown in Fig. \ref{fig1}, the source vector ${\bf{s}} = [ {{s}}_1,{{s}}_2, \ldots ,{{s}}_K]^T$ is first linearly processed by a matrix $\mathbf{U} \in \CC^{N_S \times K} $ and then transmitted over the source-relay link in the first phase. At the relay,
the received signal is first processed by $\mathbf{F}\in \CC^{N_R \times N_R}$ and then forwarded to the destination during the second phase. Vector $\mathbf{y}$ at the input of the decision device is eventually given by \cite{Rong2009LinearRelayCommunication}
\begin{equation}\label{1.1}
{\bf{y}} = {\bf{G}}\mathbf{H}\mathbf{U}{\bf{s}}+  {\bf{G}}{\bf{n}}
\end{equation}
where ${\bf{G}} \in \CC^{K
\times N_S}$ is the processing matrix used at the destination,
\begin{equation}\label{1.2}
\mathbf{H} ={\bf{H}}_{RD}{\bf{F}}{\bf{H}}_{SR}
\end{equation}
is the equivalent channel matrix, $\mathbf{H}_{SR} \in \CC^{N_R \times N_S}$ and $\mathbf{ H}_{RD}\in \CC^{N_S \times N_R}$ are the source-relay and relay-destination channel matrices, respectively. In addition, ${\bf{n}} \in \CC^{N_S \times 1}$ is a zero-mean complex Gaussian vector whose covariance matrix is $\rho\mathbf{R}_\mathbf{n}$ with $\rho>0$ {accounting for the noise variance over both links\footnote{The extension to the case in which the noise contribution over
each link has a different variance is straightforward.}} and
\begin{equation}\label{1.3}
\mathbf{R}_\mathbf{n} ={\bf{H}}_{RD}{\bf{F}}{\bf{F}}^H{\bf{H}}_{RD}^{H} + {\bf{I}}_{N_S}.
\end{equation} 
Henceforth, we denote by 
\begin{equation}\label{1.5}
{\mathbf{H}}_{SR} = {\mathbf{\Omega}_{H_{SR}}\mathbf{\Lambda}_{H_{SR}}^{1/2}\mathbf{V}_{H_{SR}}^H} 
\end{equation}
and
\begin{equation}\label{1.5.1}
{\mathbf{H}}_{RD} = {\mathbf{\Omega}_{H_{RD}}\mathbf{\Lambda}_{H_{RD}}^{1/2}\mathbf{V}_{H_{RD}}^H}
\end{equation}
the singular value decompositions (SVDs) of ${\mathbf{H}}_{SR}$ and ${\bf{H}}_{RD}$ and assume without loss of generality that the entries of the diagonal matrices $\mathbf{\Lambda}_{H_{SR}}$ and $\mathbf{\Lambda}_{H_{RD}}$ are arranged in \emph{non-increasing order}. {This amounts to saying that $\lambda_{H_{SR},k} \ge \lambda_{H_{SR},k+1}$ and $\lambda_{H_{RD},k} \ge \lambda_{H_{RD},k+1}$ for $k=1,2,\ldots,\min{(N_R, N_S)}-1$, where $\lambda_{H_{SR},k}$ and $\lambda_{H_{RD},k}$ stand for the $k$th diagonal element of ${\bf \Lambda}_{H_{SR}}$ and  ${\bf \Lambda}_{H_{RD}}$, respectively.}
Also, we denote by 

\begin{equation}
{\bf{E}} = {\rm{E}}\{
\left( {{\bf{y}} - {\bf{s}}} \right)\left( {{\bf{y}} - {\bf{s}}} \right)^H\}
\end{equation}
the mean square error (MSE) matrix. From \eqref{1.1} using \eqref{1.2} and \eqref{1.3}, it follows that
\begin{equation}\label{1.4}
{\bf{E}} = \left( {\bf{G}}\mathbf{H}\mathbf{U}-
{\bf{I}}_K \right)\left( {\bf{G}}\mathbf{H}\mathbf{U}-
{\bf{I}}_K \right)^H +\rho{\bf{G}}\mathbf{R}_\mathbf{n}{\bf{G}}^H.
\end{equation}
\vspace{-0.5 cm}
\subsection{Problem formulation}
A popular approach in the design of AF MIMO relay systems is to maximize the capacity between source and destination (see \cite{Tang2007} and \cite{Medina2007} and references therein). Although the capacity is one of the most important information-theoretic measure, there are many other ways of characterizing the reliability of transmission schemes. 
Most of them relies on the minimization/maximization of a \emph{global}
objective function $f : \RR^K \rightarrow \RR$ subject to average power constraints at the source and relay nodes. Henceforth, we assume that $f$ depends on the single MSEs $\{ [{\bf{E}}]_{k,k}; k = 1, 2 , \ldots, K\}$ and formalize such optimization problems as follows \cite{Rong2009LinearRelayCommunication} (see \cite{PalomarBook} and references therein for a detailed discussion on the subject in single-hop MIMO systems): 
\begin{equation}\label{2.1}
\mathcal P_1 : \quad \mathop {\min }\limits_{{\bf{G}},{\bf{U}},{\bf{F}}} \; f\left(\left[ {\bf{E}}\right]_{k,k};k=1,2,\ldots,K\right)\quad\quad\quad\quad\quad\quad\quad\quad\quad\quad
\end{equation}
\begin{equation}
\begin{array}{l}
\quad {\rm{s}}{\rm{.t}}{\rm{.}}\quad {\rm{tr}}\{ {\bf{U}}{\bf{U}}^H\} \le P_S \\
\quad\quad\quad \;{\rm{tr}}\left\{ {{\bf{F}} \left({\bf{H}}_{SR}{\bf{U}}{\bf{U}}^H{\bf{H}}_{SR}{^H} + \rho{\bf{I}}_{N_R}\right){\bf{F}}^H} \right\} \le P_R
 \end{array}
\end{equation}
where $P_S$ and $P_R$ denote the power available for transmission at the source and relay, respectively. Following \cite{PalomarBook}, we restrict our attention only to \textit{reasonable} $f$, i.e., functions that are increasing in each argument. 

An alternative approach is based on the minimization of the \emph{total} power consumption while meeting specific quality-of-service (QoS) requirements on the different data streams. In particular, we assume that 
the QoS constraints are given in terms of the MSEs so that the problem can be formalized as \cite{Rong2011} (see for example \cite{PalomarQoS2004} for single-hop MIMO systems):
\begin{equation}\label{2.2}
\mathcal P_2 : \quad \mathop {\min }\limits_{{\bf{G}}, {\bf{U}},{\bf{F}}} \; {\rm{tr}}\left\{ {{\bf{U}}^H {\bf{U}}}+ {{\bf{F}} \left({\bf{H}}_{SR}{\bf{U}}{\bf{U}}^H{\bf{H}}_{SR}{^H} + \rho{\bf{I}}_{N_R}\right){\bf{F}}^H} \right\}\;
\end{equation}
\begin{equation}\nonumber
\begin{array}{l}
\quad\;\,{\rm{s}}{\rm{.t}}{\rm{.}}\;\; \left[ {\bf{E}}\right]_{k,k}\le \eta_k\;\;{\rm{for}}\;\; k=1,2,\ldots,K\quad\quad\quad\quad\;
 \end{array}
\end{equation}
where $\eta_k>0$ specifies the QoS requirement for the $k$th data stream.

If not otherwise stated, in the following derivations we assume a perfect knowledge of ${\mathbf{H}}_{SR}$ and ${\mathbf{H}}_{RD}$ at all nodes.

\subsection{Design of $(\mathbf{G},{\bf{U}},{\bf{F}})$ for problem $\mathcal P_1$}
Since $f$ is increasing in each argument, the optimal $\mathbf{G}$ in (\ref{2.1}) must be such that each $\left[ {\bf{E}}\right]_{k,k}$ is minimized for any given $({\bf{U}},{\bf{F}})$ \cite{Rong2009LinearRelayCommunication}. As is well known, this is achieved by choosing $ {\bf{G}}$ equal to the Wiener filter, i.e.,
\begin{equation}\label{3.1}
 {\bf{G}}  = {\bf{U}}^H{\bf{H}}^H\left( {\mathbf{H}\mathbf{U}\mathbf{U}^H\mathbf{H}^H   +\rho{\bf{R}}_{\bf{n}}} \right)^{ - 1}.
\end{equation}
Substituting \eqref{3.1} into (\ref{1.4}) yields
\begin{equation}\label{3.2}
{\bf{E}} = {\rho}\left( \mathbf{U}^H\mathbf{H}^H{\bf{R}}_{\bf{n}}^{-1}\mathbf{H}\mathbf{U}+ {\rho}{\bf{I}}_{K} \right)^{ - 1}.
\end{equation}
It is worth observing that when the optimal $\bf{G}$ is equal to the Wiener filter the signal-to-interference noise ratio (SINR) over the $k$th stream is related to the corresponding MSE as follows 
\begin{equation}\label{3.2.1}
{\rm{SINR}}_k  = \frac{1}{[ {\bf{E}}]_{k,k}} - 1. 
\end{equation}
This means that the optimization problem in \eqref{2.1} encompasses also all design criteria in which the objective function is expressed in terms of SINRs \cite{PalomarBook}.

As shown in \cite{Rong2009LinearRelayCommunication}, the optimal ${\bf{U}}$ and ${\bf{F}}$ can be computed in closed-form for additively Schur-concave or Schur-convex functions.
As originally pointed out in \cite{PalomarBook}, this class of functions is of great interest since many different optimization criteria driving the design of wireless communication systems arise in connection with it. {The interested reader is referred to \cite{PalomarBook} for a more detailed discussion on the subject.}

\subsubsection{Additively Schur-concave functions} 
{A short list of optimization problems in which $f$ is additively Schur-concave is given below:
\begin{itemize}
\item the maximization of the mutual information $f([ {\bf{E}}]_{k,k};k=1,2,\ldots,K)= - \sum\nolimits_{k=1}^{K} \log [ {\bf{E}}]_{k,k}$;
\item the minimization of the product of the MSEs $f([ {\bf{E}}]_{k,k};k=1,2,\ldots,K)= \prod\nolimits_{k=1}^{K}[ {\bf{E}}]_{k,k}$;
\item the maximization of the sum of the SINRs $
f([ {\bf{E}}]_{k,k};k=1,2,\ldots,K)= -\sum\nolimits_{k=1}^{K}  (\frac{1}{[ {\bf{E}}]_{k,k}} - 1)$;
\item the maximization of the product of the SINRs $
f([ {\bf{E}}]_{k,k};k=1,2,\ldots,K)= -\prod\nolimits_{k=1}^{K}  (\frac{1}{[ {\bf{E}}]_{k,k}} - 1)$.
\end{itemize}}
{In writing the above list, we have used the fact that when the Wiener filter is used at the destination
the SINR is related to the MSE through \eqref{3.2.1}}.

If $f$ is additively Schur-concave then the optimal ${\mathbf{U}}$ and ${\mathbf{F}}$ in (\ref{2.1}) are given by \cite{Rong2009LinearRelayCommunication}
\begin{equation}\label{3.3}
{\mathbf{U}} ={\mathbf{\tilde V}_{H_{SR}}}{\mathbf{\Lambda}}_{U}^{1/2} 
\end{equation}
and
\begin{equation}\label{3.3.1}
{\mathbf{F}}= {\mathbf{\tilde V}_{H_{RD}}} {\mathbf{\Lambda}}_F^{1/2}\mathbf{\tilde \Omega}_{H_{SR}}^H
\end{equation}
where ${ \mathbf{\tilde V}_{H_{SR}}}$, ${\mathbf{\tilde V}_{H_{RD}}}$
and $\mathbf{\tilde \Omega}_{H_{SR}}$ are obtained from the $K$ columns
{of ${\bf V}_{H_{SR}}$, ${\bf V}_{H_{RD}}$, and ${\bf
\Omega}_{H_{SR}}$} associated to the $K$ largest singular values of the corresponding channel matrix. In addition, $\mathbf{\Lambda}_{U} \in \CC^{K\times K}$ and $\mathbf{\Lambda}_{F}\in \CC^{K\times K}$ are diagonal matrices with elements given by
\begin{equation} \label{3.4}
\lambda_{U,k} = A_{k} 
\end{equation}
and
\begin{equation} \label{3.4.1}
{\lambda_{F,k}} =  \frac{B_{k}}{A_{k}\lambda_{H_{SR},k} + \rho}
\end{equation}
where {$\lambda_{H_{SR},k}$ denotes the $k$th diagonal entry of $ \mathbf{\Lambda}_{H_{SR}}$ in \eqref{1.5}.}

The coefficients $A_{k}$ and $B_{k}$ for $k=1,2,\ldots,K$ {account for the transmission power required by the $k$th stream at the source and relay, respectively,} and are obtained as the solutions of the following problem:
\begin{equation}\label{3.5}
\mathop {\min }\limits_{\{{ A_{k} \ge 0},{ B_{k} \ge 0}\}} \; f({\boldsymbol{\lambda}})\quad\quad\quad\quad\quad\quad\quad\quad\quad\quad\quad\quad
\end{equation}
\begin{equation}\nonumber
\begin{array}{l}
\quad\quad\quad{\rm{s}}{\rm{.t}}{\rm{.}}\quad\quad \sum\limits_{k=1}^K A_{k}\le P_S \quad{\rm{and}}\quad\sum\limits_{k=1}^K B_{k}\le P_R
 \end{array}
 \end{equation}
where $\boldsymbol{\lambda} = [\lambda_1,\lambda_2,\ldots,\lambda_K]^T$ and $\lambda_{k}$ is the $k$th eigenvalue of ${\bf{E}}$. The latter is obtained substituting \eqref{3.3} and \eqref{3.3.1} into \eqref{3.2} and is given by
\begin{equation}\label{3.6}
\lambda_{k} =  \rho \frac{A_{k}\lambda_{H_{SR},k} + B_{k}\lambda_{H_{RD},k} + \rho}{\left(A_{k}\lambda_{H_{SR},k}+\rho\right)\left(B_{k}\lambda_{H_{RD},k}+\rho\right)}
\end{equation}
where {$\lambda_{H_{RD},k}$ denotes the $k$th diagonal entry of $ \mathbf{\Lambda}_{H_{RD}}$ in \eqref{1.5.1}.}

From (\ref{3.3}) and (\ref{3.3.1}), it follows that the optimal ${\mathbf{U}}$ and ${\mathbf{F}}$ match the singular vectors of the corresponding channel
matrices. In this way, the strongest spatial channels of the source-relay and relay-destination links are matched together. 

\begin{figure}[tbp]
\centering \includegraphics[width=.47\textwidth]{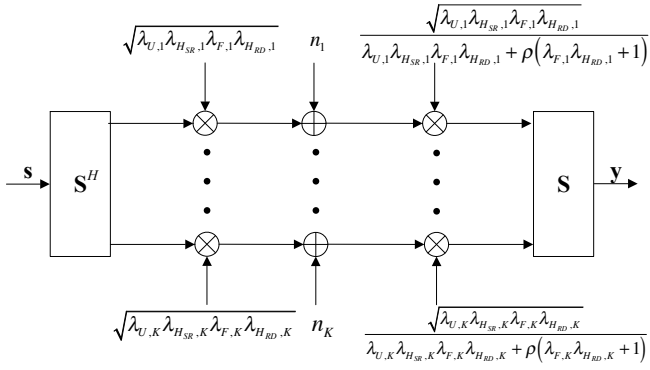}
\caption{Equivalent block diagram of the optimized one-way two-hop MIMO system when the direct link is omitted. When $f$ is additively Schur-concave, the matrix $\bf{S}$ must be chosen equal to the identity matrix. On the other hand, when $f$ is additively Schur-convex $\bf{S}$ is unitary and must be designed such that \eqref{3.52} is satisfied.}\label{fig3}
\end{figure}

Collecting the above results together, it is easily seen that the overall channel matrix ${\boldsymbol{\mathcal H}} = {\bf{G}}{\bf{H}}_{RD}{\bf{F}}{\bf{H}}_{SR} {\bf{U}}$ becomes diagonal with entries given by \cite{Rong2009LinearRelayCommunication} 
\begin{equation}\nonumber
\lambda_{{\mathcal H},k}= \frac{\lambda_{U,k}\lambda_{H_{SR},k}\lambda_{F,k}\lambda_{H_{RD},k}}{\lambda_{U,k}\lambda_{H_{SR},k}\lambda_{F,k}\lambda_{H_{RD},k} + \rho\left(\lambda_{F,k}\lambda_{H_{RD},k} + 1\right)}.
\end{equation}
{Also, the MSE matrix turns out to be diagonal with elements given by
\begin{equation}\label{3.52.1}
\left[ {\bf{E}}\right]_{k,k} =  {\lambda_{k}}.
\end{equation}}
From the above results, it follows that the AF MIMO relay system becomes equivalent to a set of parallel single-input single-output (SISO) channels. This is depicted in Fig. \ref{fig3} when ${\bf{S}}$ is chosen equal to the identity matrix. A similar result was obtained for single-hop systems \cite{PalomarBook}.

Now the only problem left is to solve \eqref{3.5}. Once $\{A_{k}\}$ and $\{B_{k}\}$ are computed, the optimal allocation of the available power over the parallel SISO channels of Fig. \ref{fig3} can be found through \eqref{3.4} and \eqref{3.4.1}. 
While in single-hop systems the optimal power distribution can easily be found by means of water-filling inspired algorithms \cite{PalomarBook}, solving \eqref{3.5} represents the most challenging task since it is not in a convex form. 
A possible solution to this problem is represented by the grid-search based algorithm illustrated in \cite{Chiang2011}. The latter was originally proposed for the maximization of the mutual information in one-way two-hop SISO multicarrier systems and has been recently extended to a generic function of the MSEs in \cite{Kim2010} {only for the simple case in which a uniform power allocation is adopted at the source.}
{Unfortunately, the algorithm in \cite{Chiang2011} is computationally intensive since a high-dense  two-dimensional grid-search whose complexity grows \emph{quadratically} with the number of subcarriers (spatial channels in the system under investigation) is required to obtain good approximations of the global minimum. To overcome this problem, the authors in \cite{Chiang2011} propose also an alternative approach based on an heuristic line of reasoning in which the power is allocated separately at the source and relay by means of a water-filling algorithm operating over a progressively searched subset of subcarriers/spatial channels. This leads to a suboptimal procedure with \emph{linear} complexity whose solution is shown in \cite{Chiang2011} to be close to the optimal one.}

An alternative approach that may guarantee a good trade-off between complexity and performance is achieved with the method proposed in \cite{Rong2009LinearRelayCommunication} and \cite{Hammerstrom2007},
where 
(\ref{3.5}) is alternately solved with respect to $\{A_{k}\}$ or $\{B_{k}\}$ keeping the other fixed. This leads to an iterative optimization procedure that if \textit{properly initialized} monotonically converges to a \textit{local} optimum of \eqref{3.5} since the conditional updates of $\{A_k\}$ and $\{B_k\}$ may either decrease or maintain (but not increase) the objective function $f({\boldsymbol{\lambda}})$. Interestingly, the minimization of $f({\boldsymbol{\lambda}})$ in \eqref{3.5} with respect to $\{A_k\}$ (or $\{B_k\}$) when $\{B_k\}$ (or $\{A_k\}$) is fixed leads to a water-filling inspired solution for most of the Schur-concave functions of interest \cite{Rong2009LinearRelayCommunication}. {However, as with any other iterative algorithms some caution must be taken in applying the above scheme since a bad initialization can prevent it from converging to a local optimum. For example, in \cite{Chiang2011} the authors observe that a uniform power allocation may not be a good initial point in fully spatial-correlated channels, i.e., $\lambda_{H_{SR},k}=\lambda_{H_{SR}}$ and $\lambda_{H_{RD},k}=\lambda_{H_{RD}}$ for all $k=1,2,\ldots,K$.}

An approximate solution of (\ref{3.5}) can be obtained in closed-form (without the need of any iterative procedure) through the low-complexity algorithm proposed in \cite{Rong2010_TCOM}. However, 
such a method works properly only when the power budgets $P_S$ and $P_R$ are sufficiently higher than the noise variance $\rho$.

\subsubsection{Additively Schur-convex functions} 
{A short list of optimization problems in which $f$ is additively Schur-convex is given below (see for example \cite{PalomarBook} and \cite{Rong2009LinearRelayCommunication} for more details):
\begin{itemize}
\item the minimization of the sum of the MSEs $f([ {\bf{E}}]_{k,k};k=1,2,\ldots,K)= \sum\nolimits_{k=1}^{K}\left[ {\bf{E}}\right]_{k,k}$;
\item the minimization of the maximum MSE $f([ {\bf{E}}]_{k,k};k=1,2,\ldots,K)= \mathop {\max }\nolimits_{1 \le k \le K} 
\left[ {\bf{E}}\right]_{k,k}$;
\item the minimization of the harmonic mean of the SINRs $f([ {\bf{E}}]_{k,k};k=1,2,\ldots,K) = \sum\nolimits_{k=1}^{K}\frac{[ {\bf{E}}]_{k,k}}{ 1 -[ {\bf{E}}]_{k,k}}$;
\item the maximization of the minimum of the SINRs $f([ {\bf{E}}]_{k,k};k=1,2,\ldots,K)= -\mathop {\min }\nolimits_{1 \le k \le K} 
\frac{ 1- [{\bf{E}}]_{k,k}}{[{\bf{E}}]_{k,k}}$.
\end{itemize}}
If $f$ is additively Schur-convex then the optimal matrices ${\mathbf{U}}$ and ${\mathbf{F}}$ in (\ref{2.1}) are given by \cite{Rong2009LinearRelayCommunication}
\begin{equation}\label{3.50}
{\mathbf{U}} ={\mathbf{\tilde V}_{H_{SR}}}{\mathbf{\Lambda}}_{U}^{1/2}\mathbf{S}^H 
\end{equation}
and
\begin{equation}\label{3.50.1}
{\mathbf{F}}= {\mathbf{\tilde V}_{H_{RD}}} {\mathbf{\Lambda}}_F^{1/2}\mathbf{\tilde \Omega}_{H_{SR}}^H
\end{equation}
where $\mathbf{S} \in \CC^{K \times K}$ is unitary while the entries of the diagonal matrices $\mathbf{\Lambda}_{U}\in \CC^{K \times K}$ and $\mathbf{\Lambda}_{F} \in \CC^{K \times K}$ are still given by \eqref{3.4} and \eqref{3.4.1}. The quantities $\{A_k\}$ and $\{B_k\}$  are now obtained as:
\begin{equation}\label{3.53}
\mathop {\min }\limits_{\{{ A_{k} \ge 0},{ B_{k} \ge 0}\}} \; \sum\limits_{k=1}^K  \rho\frac{A_{k}\lambda_{H_{SR},k} + B_{k}\lambda_{H_{RD},k} + \rho}{\left(A_{k}\lambda_{H_{SR},k}+\rho\right)\left(B_{k}\lambda_{H_{RD},k}+\rho\right)}\quad\quad\quad
\end{equation}
\begin{equation}\nonumber
\begin{array}{l}
\quad\quad{\rm{s}}{\rm{.t}}{\rm{.}}\quad \quad\sum\limits_{k=1}^K A_{k}\le P_S \quad{\rm{and}}\quad\sum\limits_{k=1}^K B_{k}\le P_R. \quad\quad\quad\quad\quad
 \end{array}
 \end{equation}
The above problem is still not convex and the same techniques illustrated previously can be applied to obtain a suboptimal solution. Interestingly, it is seen that the power allocation problem \eqref{3.53} does not depend on the particular choice of $f$.

As shown in Fig. \ref{fig3}, when $f$ is an additively Schur-convex function the optimal structure of the relay system  is diagonal up to a unitary matrix $\bf{S}$. {The latter must be chosen such that the diagonal elements of the MSE matrix ${\bf{E}}$, now given by, \cite{Rong2009LinearRelayCommunication}
\begin{equation}\label{3.54}
{\bf{E}} = {\bf{S}} {\boldsymbol{\Lambda}} {\bf{S}}^H
\end{equation}
are all equal to the arithmetic mean of its eigenvalues, i.e.,
\begin{equation}\label{3.52}
\left[ {\bf{E}}\right]_{k,k} = \frac{1}{K} \sum\limits_{i=1}^K {\lambda_{i}}
\end{equation}
where $\{{\lambda_{i}}\}$ are still of the form in \eqref{3.6}.}
If $K$ is a power of two, the above condition can easily be met choosing $\bf{S}$ equal to the discrete Fourier transform matrix or to a Walsh-Hadamard matrix. Otherwise, $\bf{S}$ can be found through the iterative procedure described in \cite{Viswanath1999}.

\begin{figure}[tbp]
\centering \includegraphics[width=.44\textwidth]{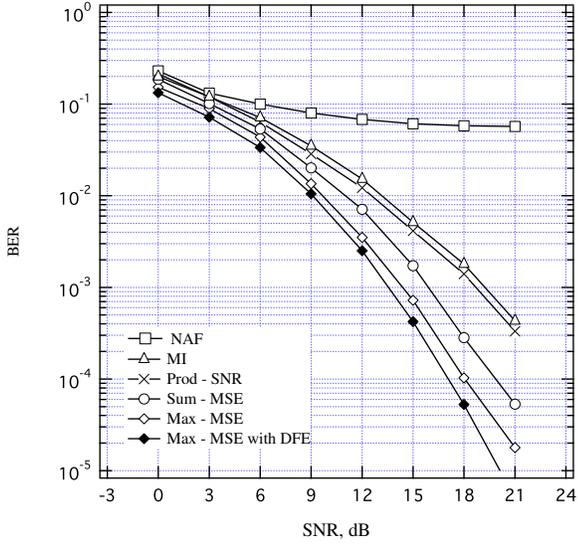}
\caption{BER of a one-way two-hop MIMO system as a function of the SNR over the source-relay link when the destination is equipped with a linear or non-linear (DFE) receiver with $N_S = N_R =3$ and $K=2$. In addition, the SNR over the relay-destination link is fixed and equal to 20 dB.}\label{fig7}
\end{figure}

Fig. \ref{fig7} illustrates the bit-error-rate (BER) of a $4-$QAM constellation as function of the signal-to-noise ratio (SNR) on the source-relay link for different optimization criteria when $N_S = N_R= 3$, $K =2$ and the SNR over the relay-destination link is fixed to 20 dB. Comparisons are made among the designs based on three Schur-concave functions: the maximization of the mutual information (MI), the maximization of the product of the SINRs (Prod - SINR) and the minimization of the sum of the MSEs (Sum - MSE); and also on a Schur-convex function\footnote{Observe that any other Schur-convex function would provide the same performance of the considered one as all of them lead to the same solution.}: the minimization of the maximum MSE (Max - MSE). The alternating algorithm developed in \cite{Rong2009LinearRelayCommunication} and \cite{Hammerstrom2007} is used to allocate the available power at source and relay. It is seen that the Schur-convex design outperforms Schur-concave ones while all the investigated solutions perform consistently better than the NAF (naive AF) design in which ${\bf{U}}$ and ${\bf{F}}$ are set equal to scaled identity matrices. {It is worth observing that MI is a good criterion only for coded systems in which the number of symbols for each coding block is large. On the other hand, the simulation setup of Fig. 3 refers to uncoded systems with a small number of symbols (4-QAM and $ K=2$) for each block and comparisons among the different schemes are made in terms of raw BER. In these circumstances, it is not surprising that the MI-based algorithm does not yield a better performance than other solutions based on different criteria.} {Moreover, the fact that the BER of a Schur-convex driven design is smaller than that of a Schur-concave one is not surprising and it is in accordance with the results illustrated in [1] for single-hop systems. A simple explanation
relies on the following observation. As shown in \eqref{3.52}, when a Schur-convex design is applied the MSEs are all equal to the arithmetic mean of the quantities $\{{\lambda_k}\}$. On the other hand, from \eqref{3.52.1} it follows that when a Schur-concave design is employed the $k$th MSE is simply equal to $\lambda_k$. Since the average BER of a MIMO system is dominated by the spatial stream with the largest MSE, it follows that a Schur-concave design cannot provide better performance than a Schur-convex one as long as the quantities $\{{\lambda_k}\}$ are different.}

\subsection{Design of $(\mathbf{G},{\bf{U}},{\bf{F}})$ for problem $\mathcal P_2$}
A close inspection of \eqref{2.2} reveals that the best we can do is to choose $\mathbf{G}$ so as to minimize each MSE \cite{PalomarQoS2004}.
This means that the optimal $\mathbf{G}$ in \eqref{2.2} is still given by the Wiener filter \eqref{3.1}.
On the other hand, the optimal $\mathbf{U}$ and $\mathbf{F}$ have the following form \cite{Rong2011}
\begin{equation}\label{4.1}
{\mathbf{U}} ={\mathbf{\tilde V}_{H_{SR}}}{\mathbf{\Lambda}}_{U}^{1/2}\mathbf{S}^H
\end{equation}
and
\begin{equation}\label{4.1.1}
{\mathbf{F}}= {\mathbf{\tilde V}_{H_{RD}}} {\mathbf{\Lambda}}_F^{1/2}\mathbf{\tilde \Omega}_{H_{SR}}^H
\end{equation}
where $\mathbf{S} \in \CC^{K \times K}$ is unitary and such that the diagonal elements of ${\bf{E}} = {\bf{S}} {\boldsymbol{\Lambda}} {\bf{S}}^H$ satisfy the following condition
\begin{equation}\label{4.2}
\left[ {\bf{E}}\right]_{k,k} = \eta_k \;\;{\rm{for}}\;\; k = 1, 2, \ldots,K.
\end{equation}
The entries of the diagonal matrices $\mathbf{\Lambda}_{U}$ and $\mathbf{\Lambda}_{F}$ are still obtained as in \eqref{3.4} and \eqref{3.4.1} with $A_{k}$ and $B_{k}$ solutions of the following problem {(the interested reader is referred to \cite{Rong2011} for more details)}:
\begin{equation}\label{4.4}
\mathop {\min }\limits_{\{A_{k} \, \ge \, 0\}, \{B_{k} \, \ge \, 0\}} \;\;\sum\limits_{k=1}^K  A_k + \sum\limits_{k=1}^K B_k \quad\quad\quad\quad\quad\quad\quad\quad\quad\quad\quad\quad\quad
\end{equation}
\begin{equation}\nonumber
\begin{array}{l}
\quad\quad{\rm{s}}{\rm{.t}}{\rm{.}}\quad  \sum \limits_{k=1}^{j} {\lambda_{k}}\le\sum \limits_{k=1}^{j}\eta_k\;\; {\rm{for}} \;\; j=1,2,\ldots,K\quad\quad\quad\quad\quad
\\
\;\;\quad\quad \quad\quad{0< \lambda_{k} \le \lambda_{k+1} \le 1}\quad k =1,2,\ldots,K-1.
 \end{array}
\end{equation}
As for the minimization of additively Schur-convex functions, the optimized structure is diagonal up to a unitary matrix $\bf{S}$. The latter must be now designed such that \eqref{4.2} is fulfilled. For this purpose, the same iterative procedure mentioned before and illustrated in \cite{Viswanath1999} can be used.

Finding the solution to \eqref{4.4} is again the major challenge of the optimization.
A possible approach is discussed in \cite{Rong2011} in which the optimal solution is upper- and lower-bounded using the geometric programming approach and the dual decomposition technique, respectively. Unfortunately, the computation complexity of both solutions is relatively high so as to make them unsuited for practical implementation. A reduced complexity algorithm is derived in \cite{Sanguinetti_Relay2011} in which the optimization in \eqref{4.4} is first carried out over $A_{k}$ and $B_{k}$ for a fixed $\lambda_{k}$ and then over all possible $\lambda_{k}$ within the feasible set of (\ref{4.4}). As shown in \cite{Sanguinetti_Relay2011}, this approach allows to 
approximate the original problem with a convex one, whose solution 
can be computed in closed-form through a multi-step procedure that requires no more than $K - 1$ steps.

\begin{figure}[tbp]
\centering \includegraphics[width=.44\textwidth]{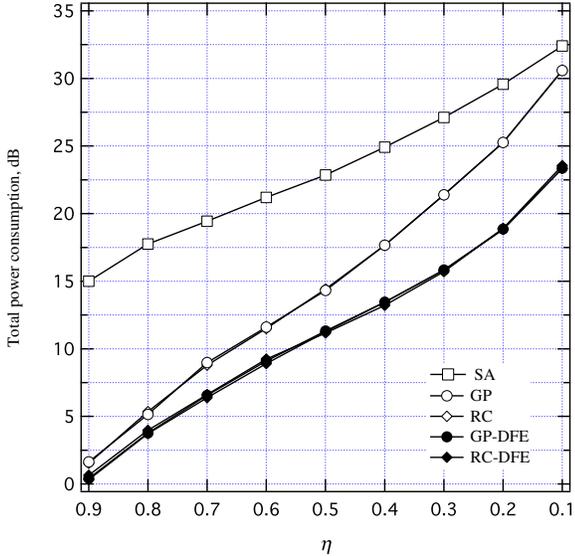}
\caption{Power consumption in dB when the destination is equipped with a linear or non-linear (DFE) receiver and $N_S = N_R = K = 3$ while equal QoS constraints are imposed, i.e., $\eta_1 = \eta_2 = \eta_3 = \eta$. In addition, the noise variances over the source-relay and relay-destination links are both fixed to $0$ dB.}\label{fig8}
\end{figure}

Fig. \ref{fig8} illustrates the total power consumption in dB when the noise variances over the source-relay and relay-destination links are both equal to $0$ dB and $N_S = N_R = K = 3 $ while the QoS constraints are for simplicity assumed to be identical, i.e., $\eta_1 = \eta_2 =\eta_3 = \eta$. The curve labelled with GP refers to a system in which the power allocation problem is approximated using the geometric programming approach proposed in \cite{Rong2011} while RC corresponds to the reduced complexity algorithm developed in \cite{Sanguinetti_Relay2011}. {Comparisons are made with SA (sub-optimal algorithm) in which the unitary matrix $\bf{S}$ in \eqref{4.1} is set equal to the identity matrix \cite{Rong2011}.} As seen, GP and RC provide substantially the same performance and achieve a remarkable gain with respect to SA. In \cite{Sanguinetti_Relay2011}, it is shown that the amount of power required by RC is very close to the minimum.

\emph{Remark.} It is worth observing that in practical applications source and relay may be unable to meet all the QoS requirements due to their limited power resource or due to regulations specifying the maximum transmit power. This calls for some countermeasures. A possible way out to this problem (not investigated yet) is represented by the technique illustrated in \cite{PalomarQoS2004} for single-hop MIMO systems in which the QoS constraints that produce the largest increase in terms of transmit power are first identified and then relaxed using a perturbation analysis. An alternative approach is to make use of an admission control algorithm such as the one illustrated in \cite{Tellambura2009} for multi-user single-antenna relay systems in which the power minimization problem is carried out jointly with the maximization of the number of users that can be QoS-guaranteed.

\subsection{Extension to non-linear architectures}

As is well-known, a decision feedback equalizer (DFE) provides a significant performance gain with respect to a linear one in single-hop MIMO systems either in terms of BER or system capacity. Similar results have recently been found also 
in one-way two-hop MIMO systems (see for example \cite{Rong2011}, \cite{Sanguinetti_Relay2011} and \cite{Rong2009NonLinearRelayCommunication}) using the same mathematical tools employed in single-hop systems. These are largely represented by the \emph{multiplicative} majorization theory and the equal-diagonal QR decomposition method illustrated
in \cite{Zhang2005}. 
As an alternative to DFE, non-linear
prefiltering based on Tomlinson-Harashima precoding (THP)
has also gained some attention
\cite{Gong2010} -- \nocite{Andrew2011}\cite{RongDual}.

Due to space limitations, we limit our attention to a DFE-based architecture for which (under the assumption of correct previous decisions) the vector $\mathbf{y}$ at the input of the decision device is given by ${\bf{y}} = \left({\bf{G}}\mathbf{H}\mathbf{U} - \mathbf{B} \right){\bf{s}}+  {\bf{G}}{\bf{n}}$ where $\mathbf{B} \in \CC^{K\times K}$ is strictly upper triangular and known as \emph{backward} matrix \cite{Rong2009NonLinearRelayCommunication}. 

\subsubsection{Design of $(\mathbf{G},{\bf{U}},{\bf{F}},{\bf{B}})$ for problem $\mathcal P_1$} 

In \cite{Rong2009NonLinearRelayCommunication}, it is demonstrated that if the objective function $f$ is \emph{multiplicatively} Schur-concave{\footnote{{{Due to space limitations, we do not report a list of multiplicatively Schur-concave or Schur-convex functions and limit to observe that they play the same role as additively Schur-concave or Schur-convex functions in the linear case. The interested reader is invited to refer to
\cite{PalomarBook} for more details.}}} 
the optimized non-linear architecture reduces to the linear one discussed previously. This means that there is no advantage in using a DFE at the destination when multiplicatively Schur-concave functions are considered. On the other hand, if $f$ is multiplicatively Schur-convex a different result is found in \cite{Rong2009NonLinearRelayCommunication}. In particular, it is shown that the optimal ${\bf{G}}$ is the Wiener filter and the optimal ${\bf{B}}$ is equal to ${\bf{B}}= {\bf{D}}{\bf{L}}^H - {\bf{I}}_{K}$ where ${\bf{L}}\in \CC^{K \times K}$ is lower triangular and such that
\begin{equation}\nonumber
{\bf{L}}{\bf{L}}^H = {\mathbf{U}^{H}\mathbf{H}^{H}\mathbf{R}_{\bf{n}}^{{-1}}\mathbf{H}\mathbf{U} +
\rho {\bf{I}}_{K}}
\end{equation}
while ${\bf{D}} \in \CC^{K \times K}$ is diagonal and designed so as to scale to unity the entries $[{\bf{D}}{\bf{L}}^{H}]_{k,k}$ for $k=1,2,\ldots,K$. The optimal ${\mathbf{U}}$ and ${\mathbf{F}}$ take the form \cite{Rong2009NonLinearRelayCommunication}
\begin{equation}\label{6.2}
{\mathbf{U}} ={\mathbf{\tilde V}_{H_{SR}}}{\mathbf{\Lambda}}_U^{1/2}\mathbf{S}^H 
\end{equation}
and
\begin{equation}\label{6.2.1}
{\mathbf{F}}= {\mathbf{\tilde V}_{H_{RD}}} {\mathbf{\Lambda}}_F^{1/2}\mathbf{\tilde \Omega}_{H_{SR}}^H
\end{equation}
where $\mathbf{S}\in \CC^{K \times K}$ is unitary while the entries of the diagonal matrices $\mathbf{\Lambda}_{U}$ and $\mathbf{\Lambda}_{F}$ are in the same form of \eqref{3.4} and \eqref{3.4.1} but with $\{A_k\}$ and $\{B_k\}$ obtained solving:
\begin{equation}\label{6.4}
\mathop {\min }\limits_{\{{ A_{k} \ge 0},{ B_{k} \ge 0}\}} \; \prod\limits_{k=1}^K  \rho\frac{A_{k}\lambda_{H_{SR},k} + B_{k}\lambda_{H_{RD},k} + \rho}{\left(A_{k}\lambda_{H_{SR},k}+\rho\right)\left(B_{k}\lambda_{H_{RD},k}+\rho\right)}\quad\quad
\end{equation}
\begin{equation}\nonumber
\begin{array}{l}
\quad\quad{\rm{s}}{\rm{.t}}{\rm{.}}\quad \quad\sum\limits_{k=1}^K A_{k}\le P_S \quad{\rm{and}}\quad\sum\limits_{k=1}^K B_{k}\le P_R. \quad\quad\quad\quad
 \end{array}
 \end{equation}
Now, the unitary matrix $\bf{S}$ is such that {(see \cite{Rong2009NonLinearRelayCommunication} for more details)}
\begin{equation}\label{6.3}
\left[{\bf{L}}\right]_{k,k}^{-1} =  \left( \prod\limits_{i=1}^K {\sqrt{\lambda_{i}}} \right)^{\frac{1}{K}}
\end{equation}
is satisfied with ${\lambda_{k}}$ defined as in \eqref{3.6}. This is achieved through the iterative algorithm illustrated in \cite{Jiang06thegeneralized}. {In addition, the MSE matrix turns out to be given by
\begin{equation}\label{6.5}
\mathbf{E} = \left(\mathbf{L}\mathbf{L}^H\right)^{-1} = \mathbf{S}\mathbf{\Lambda}\mathbf{S}^H
\end{equation}
from which using \eqref{6.3} it follows that its diagonal elements are all equal to the geometric mean of its eigenvalues \cite{Rong2009NonLinearRelayCommunication}
\begin{equation}\label{6.5}
\left[{\bf{E}}\right]_{k,k}= \left( \prod\limits_{i=1}^K {{\lambda_{i}}} \right)^{\frac{1}{K}}.
\end{equation}}

{At this stage, we observe that every increasing additively Schur-convex function is multiplicatively Schur-convex as well. Consequently, the additively Schur-convex functions analyzed in \cite{Rong2009LinearRelayCommunication} and reported in the previous section can easily be accommodated in the above framework (see \cite{PalomarBook} and \cite{D'Amico2008} for further details on this subject).} Moreover, from \eqref{6.4} it follows that, similar to the linear case, with additively Schur-convex functions the power allocation problem with multiplicatively Schur-convex functions is independent of $f$ and the optimal processing matrices lead to a channel-diagonalizing structure provided that the symbols are properly rotated at the source and destination by the unitary matrix $\bf{S}$.

Finding the globally optimal solution of \eqref{6.4} 
with algorithms of affordable complexity is extremely hard since the problem is not in a convex form. However, locally optimal solutions can be obtained resorting to the same methods illustrated previously for the linear case \cite{Rong2009NonLinearRelayCommunication}. 

{In Fig. \ref{fig7}, the curve labelled with ``Max - MSE with DFE" refers to a system in which the destination is equipped with a DFE and the design is made according to 
a multiplicatively Schur-convex function: the minimization of the maximum MSE. The available power is allocated using the alternating algorithm proposed in \cite{Rong2009LinearRelayCommunication} and \cite{Hammerstrom2007}. As expected, a non-linear system provides better performance than a linear one. This advantage is lost if a multiplicatively Schur-concave function is chosen since in these circumstances the non-linear system reduces to the linear one.}

\subsubsection{Design of $(\mathbf{G},{\bf{U}},{\bf{F}},{\bf{B}})$ for problem $\mathcal P_2$} When the power minimization problem is considered, in \cite{Rong2011} it is shown that the optimal $(\mathbf{G},\mathbf{U},\mathbf{F},\mathbf{B})$ have the same form as before for the case of multiplicatively Schur-convex functions with the only differences that $\mathbf{S}$ is such that $[{\bf{L}}]_{k,k}^{-1} = \sqrt{\eta_k}$ for $k=1,2,\ldots,K$
and the quantities $\{A_{k}\}$ and $\{B_{k}\}$ are solutions of the following problem:
\begin{equation}\label{6.20}
\mathop {\min }\limits_{\{A_{k} \, \ge \, 0\}, \{B_{k} \, \ge \, 0\}} \;\;\sum\limits_{k=1}^K  A_k + \sum\limits_{k=1}^K B_k \quad\quad\quad\quad\quad\quad\quad\quad\quad\quad\quad\quad\quad
\end{equation}
\begin{equation}\nonumber
\begin{array}{l}
\quad\quad{\rm{s}}{\rm{.t}}{\rm{.}}\quad  \prod \limits_{k=1}^{j} {\lambda_{k}}\le\prod \limits_{k=1}^{j}\eta_k\;\; {\rm{for}} \;\; j=1,2,\ldots,K\quad\quad\quad\quad\quad
\\
\;\;\quad\quad \quad\quad{0<\lambda_{k} \le \lambda_{k+1}\le 1}\quad k =1,2,\ldots,K-1.
 \end{array}
\end{equation}
Using the same arguments adopted for the linear case, in \cite{Rong2011} an upper- and a lower-bound to the globally optimal solution of the above non-convex problem are computed. Alternatively, the reduced-complexity procedure developed in \cite{Sanguinetti_Relay2011} can be used.

{In Fig. \ref{fig8}, the curves labelled with GP-DFE and RC-DFE refer respectively to a system in which  the successive GP approach of \cite{Rong2011} and the algorithm developed in \cite{Sanguinetti_Relay2011} are
employed in conjunction with a DFE. As seen, both solutions require substantially the same power and largely outperform the corresponding ones obtained with a linear receiver for all the investigated values of $\eta$.}

\subsection{Extension to frequency selective fading channels}
As done in single-hop MIMO systems, the above  optimization procedures can be extended to frequency-selective fading channels using multicarrier transmissions. To see how this comes about, assume for example that an orthogonal frequency-division multiplexing (OFDM) transmission scheme with $N$ subcarriers is used and focus on the problem of minimizing a global objective function under fixed power constraints. As illustrated in \cite{Rong2009LinearRelayCommunication} and \cite{Rong2009NonLinearRelayCommunication}, when cooperation among subcarriers is allowed, the optimization problem is formally equivalent to \eqref{2.1} both for the linear and non-linear architecture (clearly, in the latter case the optimization has also to be done with respect to the backward matrix $\mathbf{B}$, just as discussed in Section II.E). This means that if $f$ is additively (multiplicatively) Schur-concave or Schur-convex then the optimal processing matrices have the same form as before for the linear (non-linear) flat-fading case. A similar result holds true for the minimization of the total power consumption (problem $\mathcal{P}_{2})$.

Consider now the less general case in which an independent \emph{linear}  processing must be performed at each subcarrier. Denoting by ${\bf{G}}_n,{\bf{U}}_n$ and ${\bf{F}}_n$ the processing matrices operating over the $n$th subcarrier and calling ${\rm{MSE}}_{n,k}$ the MSE of the $k$th symbol over the $n$th subcarrier,
the optimization problem can be formalized as follows:
\begin{equation}\label{8.3}
\mathop {\min }\limits_{{\bf{G}}_n,{\bf{U}}_n,{\bf{F}}_n} \; f(g_1,g_2,\ldots,g_N)\quad\quad\quad\quad\quad\quad\quad\quad\quad\quad\quad\quad\quad\quad\quad\quad\quad\;
\end{equation}
\begin{equation}\nonumber
\begin{array}{l}
\quad{\rm{s}}{\rm{.t}}{\rm{.}}\quad g_n = f_n({\rm{MSE}}_{n,1}, {\rm{MSE}}_{n,2}, \ldots,{\rm{MSE}}_{n,K})\quad n=1,2,\ldots,N\\
\quad\quad\quad \; \sum\limits_{n=1}^{N}{\rm{tr}}\{ {\bf{U}}_n{\bf{U}}_n^H\} \le P_{S} \\
\quad\quad\quad \; \sum\limits_{n=1}^{N}{\rm{tr}}\left\{ {{\bf{F}}_n \left({\bf{H}}_{SR,n}{\bf{U}}_n{\bf{U}}_n^H{\bf{H}}_{SR,n}{^H} + \rho{\bf{I}}_{N_R}\right){\bf{F}}_n^H} \right\} \le P_{R}
 \end{array}
\end{equation}
where $f : \RR^N \rightarrow \RR$ and $f_n : \RR^K \rightarrow \RR$ are generic objective functions while ${\bf{H}}_{SR,n} \in \CC ^{N_R \times N_S}$ denotes the source-relay channel matrix over the $n$th subcarrier. 
Interestingly, if $f$ and $f_n$ are increasing in each argument, the above optimization problem can be greatly simplified using the primal decomposition technique which allows to decompose the given problem into $N$ independent subproblems controlled by a master problem. 
{The latter is given by
\begin{equation}\label{8.4}
\quad \mathop {\min }\limits_{\{P_{S,n} \ge 0\},\{P_{R,n} \ge 0\}} \; \hat f(\{P_{S,n}\},\{P_{R,n}\})\quad\quad\quad\quad\quad\quad\quad\quad\quad\quad\quad\quad\;
\end{equation}
\begin{equation}\nonumber
\begin{array}{l}
{\rm{s}}{\rm{.t}}{\rm{.}} \quad\sum\limits_{n=1}^{N}P_{S,n} \le P_{S} \quad{\rm{and}}\quad\sum\limits_{n=1}^{N}P_{R,n} \le P_{R} \quad\quad\quad
 \end{array}
\end{equation}
where $P_{S,n}$ and $P_{R,n}$ denote the power allocated over the $n$th subcarrier by the source and relay, respectively, while $\hat f(\{P_{S,n}\},\{P_{R,n}\}) = f(\hat g_1, \hat g_2, \ldots,\hat g_{N})$ with each $\hat g_n$ corresponding to the minimum value of the cost function $f_n$ in the following subproblem:
\begin{equation}\label{8.6}
\quad \mathop {\min }\limits_{{\bf{G}}_n,{\bf{U}}_n,{\bf{F}}_n} \; g_n = f_n({\rm{MSE}}_{n,1}, {\rm{MSE}}_{n,2}, \ldots,{\rm{MSE}}_{n,K})\quad\quad\quad\quad\quad\quad\quad
\end{equation}
\begin{equation}\nonumber
\begin{array}{l}
{\rm{s}}{\rm{.t}}{\rm{.}}
\quad {\rm{tr}}\{ {\bf{U}}_n{\bf{U}}_n^H\} \le P_{S,n} \\
\quad\quad \;\, {\rm{tr}}\left\{ {{\bf{F}}_n \left({\bf{H}}_{SR,n}{\bf{U}}_n{\bf{U}}_n^H{\bf{H}}_{SR,n}{^H} + \rho{\bf{I}}_{N_R}\right){\bf{F}}_n^H} \right\} \le P_{R,n}.
 \end{array}
\end{equation}
Using the above procedure, the solution of \eqref{8.3} can be efficiently computed as follows. The master problem in \eqref{8.4}} can be solved using the same techniques illustrated and analyzed in \cite{Palomar2005TSP} while the $n$th subproblem turns out to have the same form of \eqref{2.1} once $f$ is replaced with $f_n$. This means that when $f_n$ is additively Schur-concave or Schur-convex\footnote{Observe that in the carrier cooperative scheme the global cost function $f$ is required to be {additively} Schur-concave/convex, whereas
in the noncooperative scheme each $f_n$ must be {additively} Schur-concave/
convex.} the solution of each subproblem can be computed in closed form as shown previously for the linear flat-fading case. Such a scheme is known in the technical literature as \emph{carrier-noncooperative approach with optimal power allocation} \cite{D'Amico2008}. Alternatively, the design of the processing matrices can be performed under the assumption that no-cooperation is allowed and a fixed power (for example, a uniform distributed power) is allocated to each subcarrier. This scheme is simply referred to as a \emph{carrier-noncooperative approach} and it is again formally equivalent to \eqref{2.1}. As expected, the carrier-cooperative approach performs better than the carrier-noncooperative ones especially when highly frequency selective channels are considered \cite{Rong2009LinearRelayCommunication}. Indeed, in these circumstances the frequency diversity of the channel provides additional degrees of freedom that with cooperating subcarriers can be exploited to improve the system performance.

{The problem of minimizing the total power consumption in OFDM-MIMO systems in which several types of services are supported through spatial multiplexing has recently been investigated in \cite{Sanguinetti_2011QoS}.
Since in practical applications the reliability of each type of transmission depends on a global performance
metric measured over the assigned subcarriers, differently from \cite{Rong2011} the power minimization
problem is reformulated assuming that the QoS constraint of each service is given as a generic Schur-convex function of the MSEs over all subcarriers rather than as a set of constraints on individual MSEs. Interestingly, it turns out that the solution of
this problem reduces to the one illustrated in \cite{Rong2011} for both a linear and a non-linear architecture. The only
difference with respect to \cite{Rong2011} relies on the structure of the unitary matrix to apply to the transmitted
data symbols at the source and destination nodes.}

\subsection{Acquisition of channel state information and robust optimization}

As seen, the optimization of a one-way two-hop MIMO system requires explicit knowledge of the source-relay and relay-destination channel matrices. In principle, channel acquisition at the receiver (relay and destination) can be obtained using the same methods
employed in conventional single-hop MIMO networks (see for example \cite{Tong2004} and references therein). 
Specific algorithms for channel estimation in AF relay systems are also available in literature. For example, the estimation of both the source-relay and relay-destination channels can be performed directly at the destination node through the pilot-based schemes recently proposed in \cite{OrlikChEstTWC11} and \cite{HuaChEstTSP}. On the other hand, channel acquisition at the transmitter (source and relay) is a more demanding task. A possible solution relies on exploiting the channel reciprocity between the forward and reverse links and it is suited for \textit{open-loop} systems. Unfortunately, the reciprocity is only valid for the ``over the air'' (i.e., from antenna to antenna) segment while channel estimation is usually performed at the baseband level after the radio-frequency chain. This calls for efficient calibration schemes not easy to be implemented \cite{DiasRFImpPIRMC04}. 
An alternative approach consists in performing the estimation process at the receiver and feeding channel measurements back through a reliable reverse link. This strategy is suited for \textit{closed-loop} systems and it is nowadays considered as the most promising solution for commercial applications.

{Motivated by the above discussion, we focus on closed-loop techniques in the next. This amounts to saying that the estimation of $\mathbf{H}_{SR}$ is performed at the relay whereas the task of estimating $\mathbf{H}_{RD}$ is left to the destination node. Once estimates of  $\mathbf{H}_{SR}$ and $\mathbf{H}_{RD}$ are available, one has to decide how they should be shared among nodes for the computation of the optimal processing matrices. 
A possible solution is to make use of a \textit{distributed}\footnote{{The word ``distributed" has several meanings in wireless communications but herein it is used to refer to a system in which each node makes only use of the channel state information available locally.}} algorithm in which each node computes its own processing matrix. This means that the estimates of $\mathbf{H}_{SR}$ and $\mathbf{H}_{RD}$ must be sent from the relay to the destination and from the destination to the relay, respectively, whereas they must be both transmitted to the source node via a feedback channel from the relay. An alternative approach is represented by a \textit{centralized} algorithm in which only a single node computes all the optimal processing matrices and then transmits them to the others. Clearly, the centralized strategy requires the computation node to acquire information about all the propagation channels either by direct estimation or via feedback links. 
}

{\emph{Remark:} As mentioned in the Introduction, this work is mainly focused on transceiver design under the assumption that source and relay have perfect or at least partial knowledge of the propagation channel. As is well-known, when this assumption does not hold true and no channel state information is available during transmission, one may resort to space-time coding techniques as a means to significantly improve the link reliability and spectral efficiency of wireless communication systems. Since the pioneering work of Tarokh {\emph{et al.}} in \cite{Tarokh98}, space-time coding has gained a lot of interest both from academia and industry and a large 
number of publications has flourished in the literature in a few years (see \cite{Gesbert03} and references therein). Most of the research activity has been primarily focused on single-hop MIMO systems but it has been recently extended to relay networks with single or multiple antennas under the name of \emph{cooperative diversity} or \emph{user diversity} (see for example \cite{Wornell03}). Herein, the multiple terminals (source and/or relays) cooperate to create a virtual antenna array that provides some form of spatial diversity. Due to space limitations, we cannot provide a detailed description of the possible architectures
but we refer the interested reader to \cite{Wornell04} -- \cite{Teh11} (and references therein) for a comprehensive overview
of the literature available on this subject. 
}}

\begin{figure*}[tbp]
\centering \includegraphics[width=0.97\textwidth]{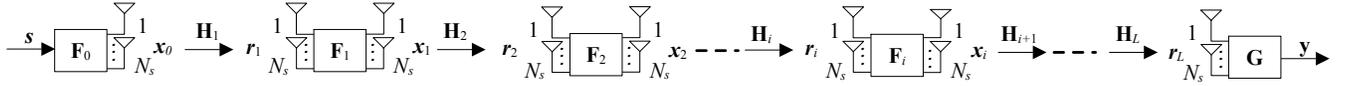}
\caption{Block diagram of one-way multi-hop MIMO systems.}\label{fig6}
\end{figure*}

\subsubsection{Robust Linear Optimization}

{Assume that estimates $\hat{\mathbf{H}}_{SR}$ and $\hat{\mathbf{H}}_{RD}$ of the source-relay and relay-destination channels are available.
The simplest solution is to use them in place of $\mathbf{H}_{SR}$ and $\mathbf{H}_{RD}$ for the computation of the optimal processing matrices.} Although simple, such an approach leads to a substantial performance degradation with respect to the perfect knowledge case.
{An alternative route consists in designing the optimal matrices taking the channel estimation errors into account. This leads to the so-called \textit{robust} optimization design in which either min-max approaches or stochastic methods are usually employed \cite{PalomarMaxminTSP06}. In the sequel, we concentrate on stochastic methods and review some recent works developed specifically for a linear AF architecture. 
}

{The robust design of $({\mathbf{G}},{\mathbf{F}})$ that minimizes the sum of the MSEs with a power constraint at the relay has been addressed in \cite{XingRobustTSP10}. The main difference with respect to the perfect channel knowledge case is that all the statistical expectations involved in the optimization problem are computed also with respect to the distribution of the channel estimation errors. To this end, in \cite{XingRobustTSP10} the following error models are assumed (see also \cite{MusavianTVT07} and \cite{DingMmseCeTSP09} and references therein):
\begin{equation}\label{SEM}
    \mathbf{H}_{SR}=\hat{\mathbf{H}}_{SR}+ \boldsymbol{\Delta}_{SR} \end{equation}
    and
\begin{equation}\label{SEM.1}
\mathbf{H}_{RD}= \hat{\mathbf{H}}_{RD}+ \boldsymbol{\Delta}_{RD} 
\end{equation}
where 
\begin{equation}\label{SEM1}
\boldsymbol{\Delta}_{SR}=\boldsymbol{\Sigma}_{SR}^{1/2}\boldsymbol{\Delta}_{SR}^{(w)}\boldsymbol{\Psi}_{SR}^{1/2} 
\end{equation}
and
\begin{equation}\label{SEM1.1}
\boldsymbol{\Delta}_{RD}=
\boldsymbol{\Sigma}_{RD}^{1/2}\boldsymbol{\Delta}_{RD}^{(w)}\boldsymbol{\Psi}_{RD}^{1/2}
\end{equation}
are the channel estimation error matrices. In particular, the elements of $\boldsymbol{\Delta}_{SR}^{(w)} \in \mathbb{C}^{N_{R} \times N_{S}}$ $\big(\boldsymbol{\Delta}_{RD}^{(w)} \in \mathbb{C}^{N_{S} \times N_{R}} \big)$ are independent and identically distributed zero-mean complex Gaussian random variables with unit variance, while $\boldsymbol{\Sigma}_{SR} \in \mathbb{C}^{N_{R} \times N_{R}}$ $\big(\boldsymbol{\Sigma}_{RD} \in \mathbb{C}^{N_{S} \times N_{S}}\big)$ and $\boldsymbol{\Psi}_{SR}^{T} \in \mathbb{C}^{N_{S} \times N_{S}}$ $\big(\boldsymbol{\Psi}_{RD}^{T} \in \mathbb{C}^{N_{R} \times N_{R}}\big)$ are the row and column covariance matrices of $\boldsymbol{\Delta}_{SR}$ ($\boldsymbol{\Delta}_{RD}$), respectively.
Assuming that the covariance matrices are perfectly known and that $\hat{\mathbf{H}}_{SR}$ and $\hat{\mathbf{H}}_{RD}$ are given, 
the channel uncertainties in \eqref{SEM} and \eqref{SEM.1} are only represented by $\boldsymbol{\Delta}_{SR}^{(w)}$ and $\boldsymbol{\Delta}_{RD}^{(w)}$ in \eqref{SEM1} and \eqref{SEM1.1}. In these circumstances, it is easily recognized that
the \textit{a posteriori} distributions of the matrices $\mathbf{H}_{SR}$ and $\mathbf{H}_{RD}$ in \eqref{SEM} and \eqref{SEM.1} follow the well-known Gaussian-Kronecker model \cite{XingRobustTSP10}. }

It is worth observing that the expressions of $\boldsymbol{\Sigma}_{SR}$, $\boldsymbol{\Psi}_{SR}$, $\boldsymbol{\Sigma}_{RD}$ and $\boldsymbol{\Psi}_{RD}$ depend on the specific channel estimation algorithm. For example, it can easily be shown \cite[Remark 1]{XingRobustTSP10} that the above model includes two well-known Bayesian MMSE estimators proposed in literature \cite{MusavianTVT07} and \cite{DingMmseCeTSP09}. 

{The robust design of $({\mathbf{G}}, {\mathbf{U}},{\mathbf{F}})$ has been addressed in \cite{ChaliseImperfectCSIJASP10} for a flat-fading scenario and in \cite{XingRobustOFDMTSP10} for an OFDM system operating over a frequency-selective fading channel. The extension of these results to additively Schur-concave and Schur-convex functions can be found in \cite{RongRobustTSP11} and is now briefly reviewed.} 

{We start observing that the optimization problem in \eqref{2.1}, when all nodes have imperfect channel state information, takes the following form \cite{RongRobustTSP11}:
\begin{equation}\label{RDOP}
\bar{\mathcal P_{1}} : \quad \mathop {\min }\limits_{{\bf{G}},{\bf{U}},{\bf{F}}} \; f\left(\left[ {\overline{\bf{E}}}\right]_{k,k};k=1,2,\ldots,K\right)\quad\quad\quad\quad\quad\quad\quad\quad\quad\quad\quad\quad\quad\;
\end{equation}
\begin{equation}\nonumber
\begin{array}{l}
\quad {\rm{s}}{\rm{.t}}{\rm{.}}\quad {\rm{tr}}\left\{ {\bf{U}}{\bf{U}}^H\right\} \le P_S \\
\quad\quad\quad \;{\rm{tr}}\left\{ {{\bf{F}} \left({\hat{\bf{H}}}_{SR}{\bf{U}}{\bf{U}}^H{\hat{\bf{H}}}_{SR}^H+ \alpha\boldsymbol{\Sigma}_{SR}+\rho{\bf{I}}_{N_{R}}\right){\bf{F}}^H} \right\} \le P_R
 \end{array}
\end{equation}
where $\alpha=\mathrm{tr}\{\mathbf{U}\mathbf{U}^{H}\boldsymbol{\Psi}_{SR}\}$ and $\overline{\bf{E}}=\mathrm{E}_{\mathbf{H}_{SR},\mathbf{H}_{RD}}\{\mathbf{E}\}$ denotes the expectation of the MSE matrix in \eqref{1.4} computed with respect to the statistical distributions of $\mathbf{H}_{SR}$ and $\mathbf{H}_{RD}$ in \eqref{SEM} and \eqref{SEM.1}. This yields
%
\begin{equation}
\label{AverE}\nonumber
\overline{\bf{E}}=\mathbf{G}\mathbf{A}\mathbf{G}^{H}-\mathbf{G}\mathbf{\hat H} \mathbf{U} - \mathbf{U}^{H}\mathbf{\hat H}^H \mathbf{G}^H+\mathbf{I}_{K}
\end{equation}
where we have defined $\mathbf{\hat H} = {\hat{\bf{H}}}_{RD} \mathbf{F} {\hat{\bf{H}}}_{SR}$ and 
\begin{equation}
\label{ }
\nonumber
\mathbf{A}=\beta \hat{\mathbf{H}}_{RD}\mathbf{F}\left(\hat{\mathbf{H}}_{SR}\mathbf{U}\mathbf{U}^{H}\hat{\mathbf{H}}_{SR}^{H}+\alpha \boldsymbol{\Sigma}_{SR} + \rho{\bf{I}}_{N_{R}}\right) {\bf{F}}^{H}\hat{\mathbf{H}}_{RD}^{H}\boldsymbol{\Sigma}_{RD}+{\bf{I}}_{N_{S}}
\end{equation}
with 
\begin{equation}
\label{ }
\nonumber
\beta={\rm{tr}}\{{{\bf{F}} ({\hat{\bf{H}}}_{SR}{\bf{U}}{\bf{U}}^H{\hat{\bf{H}}}_{SR}^H+ \alpha\boldsymbol{\Sigma}_{SR}+\rho{\bf{I}}_{N_{R}}){\bf{F}}^H} \boldsymbol{\Psi}_{RD}\}.
\end{equation}
The optimal $\mathbf{G}$ in \eqref{RDOP} is the Wiener filter given by $\mathbf{G}=\mathbf{U}^H\mathbf{\hat H}^H\mathbf{A}^{-1}$.
On the other hand, finding the optimal $\mathbf{U}$ and $\mathbf{F}$ for arbitrary covariance matrices is very difficult. 
In \cite{RongRobustTSP11} their explicit structure is provided only when the row and/or the column covariance matrices are equal to scaled identity matrices.
For example, assume that $\boldsymbol{\Sigma}_{SR}=\epsilon_{SR}\mathbf{I}_{N_{R}}$ and $\boldsymbol{\Sigma}_{RD}=\epsilon_{RD}\mathbf{I}_{N_{S}}$ and let the SVDs of $\hat{\mathbf{H}}_{SR}$ and $\hat{\mathbf{H}}_{RD}$ be respectively given by 
\begin{equation}\nonumber
\hat{\mathbf{H}}_{SR}={{\mathbf{ \Omega}}_{\hat H_{SR}}{\mathbf{\Lambda}}_{\hat H_{SR}}^{1/2}{\mathbf{V}}_{\hat H_{SR}}^H}\end{equation}  
and 
\begin{equation}\nonumber
\hat{\mathbf{H}}_{RD}={{\mathbf{\Omega}}_{\hat H_{RD}}{\mathbf{\Lambda}}_{\hat H_{RD}}^{1/2}{\mathbf{V}}_{\hat H_{RD}}^H}.
\end{equation}
In the above circumstances, if additively Schur-concave functions are considered the optimal $\mathbf{U}$ and $\mathbf{F}$ take the form \cite{RongRobustTSP11}
\begin{equation}
\label{RD-optUF}
\mathbf{U}=\widetilde{\mathbf{V}}_{\hat H_{SR}} \boldsymbol{\Lambda}_{U}^{1/2} 
\end{equation} 
and
\begin{equation}
\label{RD-optUF.1}
\mathbf{F}=\widetilde{\mathbf{V}}_{\hat H_{RD}} \boldsymbol{\Lambda}_{F}^{1/2}\widetilde{\mathbf{\Omega}}_{\hat H_{SR}}^{H}
\end{equation} 
where $\widetilde{\mathbf{\Omega}}_{\hat H_{SR}}$, $\widetilde{\mathbf{V}}_{\hat H_{SR}}$ and $\widetilde{\mathbf{V}}_{\hat H_{RD}}$ correspond to the $K$ columns of ${\mathbf {\Omega}}_{\hat H_{SR}}$, ${\mathbf{V}}_{\hat H_{SR}}$ and ${\mathbf{V}}_{\hat H_{RD}}$ associated to the $K$ largest singular values. The entries of the diagonal matrices $\boldsymbol{\Lambda}_{U}\in \CC^{K \times K}$ and $\boldsymbol{\Lambda}_{F} \in \CC^{K \times K}$ can be obtained through the iterative method developed in \cite{Rong2009LinearRelayCommunication}. If $f$ is a Schur-convex function, the optimal $\mathbf{F}$ is as in \eqref{RD-optUF.1} while the optimal $\mathbf{U}$ is given by 
\begin{equation}
\mathbf{U}=\widetilde{\mathbf{V}}_{\hat H_{SR}} \boldsymbol{\Lambda}_{U}^{1/2} \mathbf{S}^H
\end{equation} 
with $\mathbf{S}$ being a proper unitary matrix chosen such that the diagonal elements of $\overline{\bf{E}}$ are all equal to the arithmetic mean of its eigenvalues.}

{As mentioned before, in \cite{XingRobustTSP10} and \cite{ChaliseImperfectCSIJASP10} -- \cite{RongRobustTSP11} it is shown that a robust design provides better performance compared to the simple scheme in which the channel estimation errors are not taken into account and the estimated channel matrices $\hat{\mathbf{H}}_{SR}$ and $\hat{\mathbf{H}}_{RD}$ are simply used in place of $\mathbf{H}_{SR}$ and $\mathbf{H}_{RD}$.}

\section{Optimization of a one-way multi-hop MIMO system}

{In case of a long source-destination distance, multi-hop communications may be necessary to carry the information from the source to the destination. Consider for example a linear $L-$hop MIMO
relay system consisting of $L-1$ relays each equipped (for notational simplicity) with the same number $N_S$ of antennas employed at source and destination nodes and call ${\bf H}_{i} \in \CC^{N_S \times N_S} $ the channel matrix between the $i$th and
the $(i-1)$th nodes or hops (see Fig. \ref{fig6}). Also, denote by $\mathbf{F}_0 \in \CC^{N_S \times K}$ and ${\bf F}_{i}\in \CC^{N_S \times N_S}$ for $i=1,2,\ldots,L-1$ the source and the $i$th relay processing matrices, respectively. In these circumstances, the vector received at the $i$th relay takes the form 
\begin{equation}
{\bf r}_i= {\bf H}_{i}{\bf x}_{i-1}+ {\bf n}_i, \qquad
i=1,2,\ldots,L \label{yi}
\end{equation}
where $\mathbf{x}_0 = \mathbf{F}_0\mathbf{s}$ with $\mathbf{s} \in \CC^{K \times 1}$and ${\bf x}_{i} \in \CC^{N_S \times 1}$ for $i=1,2,\ldots,L-1$ is the signal
vector transmitted by the $i$th relay given by 
\begin{equation}\label{xi}
{\bf x}_i = {\bf F}_{i} {\bf
r}_i
\end{equation}
while ${\bf n}_i \in \CC^{N_s \times 1}$ accounts for thermal noise. 

{Some recent works on the multi-hop MIMO systems
described by the above model can be found in
\cite{Yeh} -- \cite{Fawaz}. In particular, in \cite{Yeh} the asymptotic capacity is derived under the assumption that ${\bf F}_{i}$ for $i=1,2,\ldots,L-1$ is a scaled identity matrix while the capacity
scaling law with an
asymptotically large number of hops is computed in \cite{Wagner2}.
In \cite{Yang} the authors investigate the achievable diversity gain when diagonal
relaying matrices are used. In \cite{Fawaz} the optimal $\{{\bf F}_{i}\}$ are found by neglecting the noise at
the relay nodes.}

{The first attempt at designing all the involved processing matrices according to a different optimization criterion can be found in
\cite{RongMH} in which the minimization of
additively Schur-concave/convex objective functions
subject to average power constraints at the source and the relays is considered. Although conceptually similar to the one discussed in Section II, the above problem is much more involved since the objective function now depends on all the relay amplifying matrices and the power constraint at each relay node is a function of the processing matrices of all backward nodes.
Interestingly, it turns out that the solution of such a complicated problem has the same form of the channel-diagonalizing structure found before for a two-hop system. Furthermore, this elegant result is valid for any arbitrary number $L$ of hops. Mathematically, $\bf{G}$ is the Wiener filter while the optimal ${{\bf F}}_0$ and ${{\bf F}}_i$ for $i=1,2,\ldots,L -1$ are given by \cite{RongMH}
\begin{equation}
{\mathbf F_0} = {{\mathbf {\tilde V}}}_{H_1}{\bf
\Lambda}_{U}^{1/2}{\bf S}^H 
\end{equation}
and
\begin{equation}
{\bf F}_{i} =
{\mathbf {\tilde V}}_{H_{i+1}} {\bf \Lambda}^{1/2}_{F_{i}} {\bf
\mathbf {\tilde \Omega}}^H_{H_{i}}\label{F}
\end{equation}
where ${\bf S}\in \CC ^{K \times K}$ is unitary while ${\mathbf {\tilde \Omega}}_{H_i} \in \CC^{N_S \times K}$ and ${\mathbf {\tilde V}}_{H_i}\in \CC^{N_S \times K}$ are obtained from the SVD of ${\bf H}_i$. In addition, $\mathbf{\Lambda}_{U}\in \CC^{K \times K}$ and $\{{\bf \Lambda}_{F_{i}}\in \CC^{K \times K}\}$ are
the diagonal power loading matrices that can be designed for example using the alternating power loading algorithm described in Section II. As for two-hop systems, ${\bf S}$ is equal to the identity matrix when
additively Schur-concave functions are considered while it must be
such that the overall MSE matrix has identical diagonal elements for additively
Schur-convex functions. All the above results have been later extended to non-linear architectures in
\cite{Rong2009NonLinearRelayCommunication} while the minimization of
the power consumption with QoS requirements is discussed in
\cite{Rong2011}.}

{Although interesting from a mathematical point of view, the realization of the above optimized multi-hop system is a challenging task since
centralized processing is required to compute the optimal ${\bf\Lambda}_{U}$ and $\{{\bf\Lambda}_{F_i}\}$. This may lead to a system with high
computational complexity and large signaling overhead. To
overcome this difficulty, simplified algorithms are proposed in
\cite{RongMURelay} in which the optimization of the relay matrices is carried out locally at each relay node while maintaining comparable performance with respect to the optimal one. This is enabled by the
observation that the optimal processing matrix of each relay can be rewritten
as the combination of two linear filters that allow to decompose the overall MSE matrix ${\bf E}$ into the sum of the MSE
matrices at all relays. 

{When the instantaneous channel state information
is only available at the destination but it is unknown at the source and the relays, the structure of the optimal source and
relay amplifying matrices that maximize the source-destination ergodic sum capacity is derived in \cite{RongNoCSI} using knowledge of the channel covariance matrices. 
}

\emph{Remark.} When the source is far away from the destination and  a large number of hops is required, the noise propagation effect of non-regenerative relays makes data recovery at the destination almost impossible for practical values of SNRs. In these circumstances, a combination of regenerative and non-regenerative relays should be used to provide a good tradeoff between the end-to-end processing delay and error rate performance \cite{RongMH}.

\begin{figure}[tbp]
\centering \includegraphics[width=.47\textwidth]{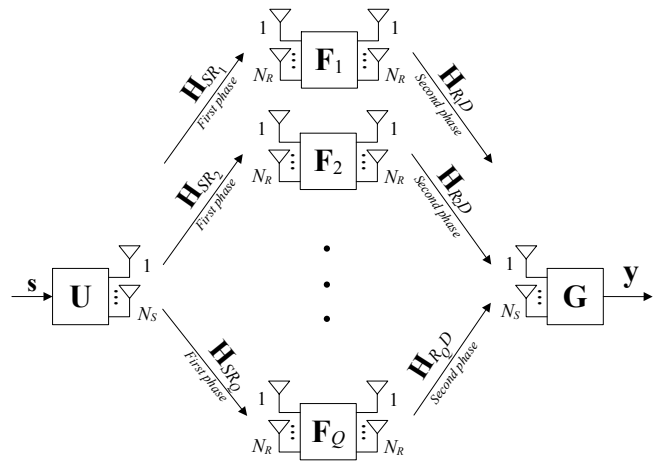}
\caption{Block diagram of one-way two-hop linear MIMO systems with multiple relays.}\label{fig5}
\end{figure}

\section{Optimization of a one-way two-hop MIMO system with multiple (parallel) relays}
The block diagram of a linear one-way two-hop MIMO system with multiple (parallel) relays is shown in Fig. \ref{fig5} in which the number of active relays is denoted by $Q$ and the matrix collecting the channel gains between the source and the $q$th relay (the $q$th relay and the destination) for $q=1,2,\ldots, Q$ is called $\mathbf H_{SR_q }\in \CC^{N_R \times N_S}$ ($\mathbf H_{R_q D }\in \CC^{N_S \times N_R}$). In these circumstances, the vector ${\bf{y}}\in \CC^{K \times 1}$ can be expressed as in \eqref{1.1} with the only difference that the matrices $\mathbf H_{SR}\in \CC^{N_RQ \times N_S}$ and $\mathbf H_{RD}\in \CC^{ N_S \times N_RQ }$ now take the form $\mathbf H_{SR}  = [ {\mathbf H_{SR_1 }^H \,\mathbf H_{SR_2 }^H \,\cdots \,\mathbf H_{SR_Q }^H }]^H$ and $\mathbf H_{RD}  = [ {\mathbf H_{R_1 D}\,\mathbf H_{R_2 D} \, \cdots \, \mathbf H_{R_Q D} }]$ while ${\bf{F}} \in \CC^{N_RQ \times N_RQ}$ is block diagonal and given by ${\bf{F}} = {\rm{diag}}\left\{ {{\bf{F}}_1 ,{\bf{F}}_2 , \ldots ,{\bf{F}}_Q }\right\}$.

In \cite{Eltawil2008} the authors set $\bf{G}$ and $\bf{U}$ equal to the identity matrix and attempt to find the optimal structure of $\mathbf {F}$ minimizing the sum of the MSEs subject to a \emph{global power} constraint at the output of the relays. Unfortunately, the solution is found in closed-form only for the simple case in which relays are equipped with a single antenna, i.e., $N_R = 1$ while the multiple antenna case is addressed without imposing any power constraint at the relays. The design of $\mathbf {F}$ that minimizes the total power consumption while fulfilling a given set of SNR constraints is investigated in \cite{Luo2008} and a power efficient solution is derived in closed-form after solving a two-step optimization problem. 
The extension to a system in which $\bf{U}$ and $\bf{G}$ may have a general structure has recently been considered in \cite{Rong2010-LinearMultipleRelay} for the minimization of the sum of the MSEs. In particular, in \cite{Rong2010-LinearMultipleRelay} it is found that the optimal $\bf{G}$ is the Wiener filter while the optimal $\bf{U}$ and $\bf{F}$ are such that
\begin{equation}\label{11.5}
{\mathbf{U}} ={\mathbf{\tilde V}_{H_{SR}}}{\mathbf{\Lambda}}_{U}^{1/2} 
\end{equation}
and
\begin{equation}\label{11.5.1}
{\mathbf{H}_{RD}}{\mathbf{F}}= \mathbf{P}{\mathbf{\Lambda}}_F^{1/2}\mathbf{\tilde \Omega}_{H_{SR}}^H.
\end{equation}
In the above equations, $ \mathbf{P} \in \CC^{K \times K}$ is an arbitrary unitary matrix while ${\mathbf{\tilde \Omega}_{H_{SR}}} \in \CC^{QN_R\times K}$ and ${\mathbf{\tilde V}_{H_{SR}}} \in \CC^{N_S\times K}$ are obtained from the SVD of $\mathbf H_{SR}$. From the above results, it follows that the diagonalization of the overall channel matrix ${\boldsymbol{\mathcal H}} = {\mathbf{H}_{RD}}{\mathbf{F}} \mathbf{H}_{SR}\mathbf{U}$ is achieved up to a unitary matrix $\bf{P}$. Unfortunately, in \cite{Rong2010-LinearMultipleRelay} such a matrix is arbitrarily set equal to the identity since finding its optimal structure is an extremely hard problem whose solution is not known yet. On the other hand, the evaluation of the elements of the diagonal matrices ${\mathbf{\Lambda}}_{U}$ and ${\mathbf{\Lambda}}_{F}$ requires to solve a non-convex power allocation problem which can be closely approximated using the same arguments outlined in Section II.

\begin{figure}[tbp]
\centering \includegraphics[width=.44\textwidth]{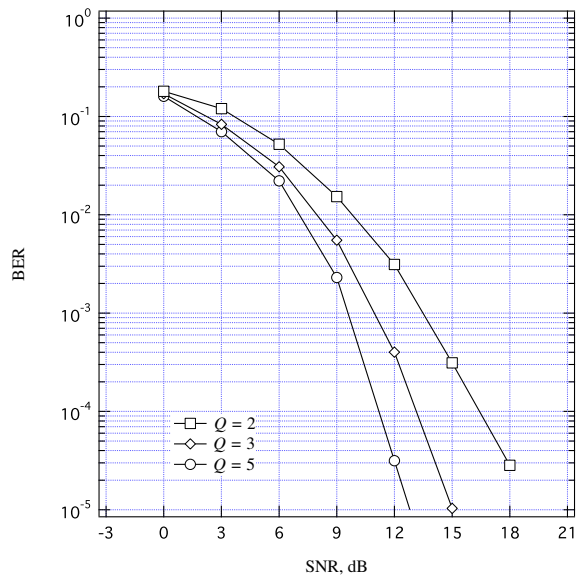}
\caption{BER of a one-way two-hop MIMO system as a function of the SNR over the source-relay links when $Q= 2,3$ and $5$ multiple relays are used and $N_S = N_R = K = 3$. In addition, the SNRs over the relay-destination links are fixed to 20 dB.}\label{fig9}
\end{figure}

Fig. \ref{fig9} illustrates the BER of the system designed in \cite{Rong2010-LinearMultipleRelay} as a function of the SNR over the source-relay links when the symbols belong to a $4-$QAM constellation and $Q$ varies from $2$ to $5$. In addition, the SNRs over the relay-destination links are fixed to 20 dB. As expected, increasing the number of relays improves the system performance. Numerical results not shown for space limitations show that a substantial improvement is achieved with respect to {\cite{Eltawil2008}} in which only the processing matrix ${\bf{F}}$ is optimized. 

The above results have recently been extended in \cite{Rong2010-NonLinearMultipleRelay} to a system in which a DFE is used at the destination. In particular, it is found that the channel-diagonalizing structure is optimal provided that two unitary matrices are used at the source (destination) to rotate (counter-rotate) the transmitted (received) symbols. As before, finding the optimal structure of the above unitary matrices is an open issue.

The optimization of $(\mathbf{G,F})$ for minimizing the power consumption subject to MSE-based QoS constraints is investigated in \cite{Fu2011} with either an optimal or non-optimal source precoding matrix $\mathbf{U}$.

\section{Optimization of a one-way two-hop MIMO system in the presence of direct link}\label{DirectLink}

{When the direct link is sufficiently strong, the exchange of information between source and destination still requires four phases as in the case in which the direct link is negligible. However, differently from this latter case both destination and relay receive the source signal in the first phase. On the other hand, during the second phase the source node is silent and the relay node sends the linearly precoded signal to the destination. In these circumstances,} the vector $\mathbf{y}$ turns out to be in the same form of \eqref{1.1} with the equivalent channel matrix ${\bf{H}} = {{\bf{H}}_{RD}{\bf{F}}{\bf{H}}_{SR}}$ replaced by ${\bf{H}}^{(DL)} = [{\bf{H}}^H_{SD}, \;{{\bf{H}}^H }]^H$, where $\mathbf{H}_{SD} \in \CC^{N_S \times N_S}$ is the source-destination channel matrix. In addition, the covariance matrix of the noise vector $\bf{n}$ is found to be $\rho\mathbf{R}^{(DL)}_\mathbf{n}$ with
\begin{equation}\nonumber\label{9.3}
{\bf{R}}^{(DL)}_{\bf{n}}  = \left[ {\begin{array}{*{20}c}
   {\bf{I}}_{N_S} & {\bf{0}}_{N_S}  \\
   {\bf{0}}_{N_S} & \mathbf{R}_\mathbf{n} \\
\end{array}} \right]
\end{equation}
and $\mathbf{R}_\mathbf{n}$ given by \eqref{1.3}.

Although there exists much ongoing research, the problem of jointly designing $({\bf{G}},{\bf{U}},{\bf{F}})$ when the direct link is present is still much open. 
A first attempt 
can be found in \cite{Tang2007} in which the maximization of the mutual information is considered and the design of $\bf{F}$ is carried out under the assumption that no operation is performed at the source and destination. This amounts to setting $\bf{G}$ and $\bf{U}$ equal to the identity matrix. Unfortunately, finding the solution of the above problem turns out to be extremely challenging and only upper- and lower-bounds of the mutual information are used to compute $\bf{F}$. The same problem is considered in \cite{Medina2007} wherein a suboptimal structure of $\bf{F}$ is derived following a different line which does not take the power constraint at the relay into account. In \cite{Rong2009_CommLetter}, $\bf{U}$ is chosen equal to the identity matrix and the joint design of $(\bf{G},\bf{F})$ is based on the following optimization problem
\begin{equation}\label{9.10}
\quad \mathop {\min }\limits_{{\bf{G}},{\bf{F}}} \; f\left(\left[ {\bf{E}}\right]_{k,k};k=1,2,\ldots,K\right)\quad\quad\quad\quad\quad\quad\quad\quad\quad\;
\end{equation}
\begin{equation}\nonumber
\begin{array}{l}
{\rm{s}}{\rm{.t}}{\rm{.}}\quad \;{\rm{tr}}\left\{ {{\bf{F}} \left({\bf{H}}_{SR}{\bf{H}}^H_{SR} + \rho{\bf{I}}_{N_R}\right){\bf{F}}^H} \right\} \le P_R
 \end{array}
\end{equation}
where $f$ is a generic increasing function of its arguments. As shown in \cite{Rong2009_CommLetter}, the optimal ${\bf{G}}$ is again  the Wiener filter
while the relay processing matrix ${\bf{F}}$ is found solving \eqref{9.10}
with ${\bf{E}}$ now given by
\begin{equation}\nonumber\label{6.5}
{\bf{E}} = {\rho}\left( \mathbf{H}_{SD}^H\mathbf{H}_{SD} + \mathbf{H}^H{\bf{R}}_{\bf{n}}^{-1}\mathbf{H}+ {\rho}{\bf{I}}_{K} \right)^{ - 1}.
\end{equation}
From the right-hand-side of the above equation, it is seen that differently from \eqref{3.2} an additional term $\mathbf{H}_{SD}^H\mathbf{H}_{SD}$ accounting for the direct link is present in the MSE matrix.
The optimal $ {\bf{F}}$ in \eqref{9.10} has the following form \cite{Rong2009_CommLetter}
 \begin{equation}\nonumber\label{9.12}
{\mathbf{F}}= \tilde{\mathbf{V}}_{H_{RD}} {\mathbf{A}}\tilde{\mathbf{\Omega}}_{H_{SR}}^H
\end{equation}
where $\mathbf{A} \in \CC^{K \times K}$ is an arbitrary matrix. The above result indicates that ${\mathbf{F}}$ is a general linear beamforming matrix matched to the left (right) singular vectors of the source-relay (relay-destination) channel. The only difference with respect to the case in which the direct link is neglected is that $\mathbf{A}$ is not diagonal unlike ${\mathbf{\Lambda}}_F$ in \eqref{3.3.1}. This means that the optimized structure does not lead to a diagonalization of the overall communication system. In \cite{Rong2009_CommLetter} a closed form solution for the optimal $\mathbf{A}$ is provided only for the simple case in which a single antenna is employed at the source. On the other hand, when multiple antennas are employed $\mathbf{A}$ is optimized only by means of numerical methods.

In \cite{Tseng2010} the authors deal with the problem of designing $(\bf{G},\bf{U},\bf{F})$ so as to minimize the sum of the MSEs subject to power constraints at the source and relay. As expected, the optimal ${\bf{G}}$ is found to be the Wiener filter while the joint design of ${\bf{U}}$ and ${\bf{F}}$
is addressed by means of a suboptimal two-stage procedure not based on any optimality criterion.  
The same problem is addressed in \cite{Rong2010_TCOM} where it is shown that the optimal $\bf{F}$ has the following structure:
 \begin{equation}\nonumber\label{6.6}
{\mathbf{F}}= \tilde{\mathbf{V}}_{H_{RD}} {\mathbf{A}}\tilde{\mathbf{\Omega}}^H
\end{equation}
where ${\mathbf{A}}\in \CC^{K \times K}$ is an arbitrary matrix while $\tilde {\mathbf{\Omega}} \in \CC^{N_s \times K}$ is obtained from the SVD of ${\bf{H}}_{SR}{\bf{U}} = \mathbf{\Omega}\mathbf{\Lambda}\mathbf{V}^H$ and corresponds to the $K$ columns of $\mathbf{\Omega}$ associated to the $K$ largest singular values.
As seen, ${\mathbf{F}}$ is given as a function of $\bf{U}$ and ${\mathbf{A}}$. Unfortunately, in \cite{Rong2010_TCOM} no closed-form is provided for the joint design of ${\bf{U}}$ and ${\mathbf{A}}$. Only a suboptimal solution making use of gradient-based numerical methods is discussed. Fig. \ref{fig10} illustrates the BER of the algorithm developed in \cite{Rong2010_TCOM} as a function of the SNR over the source-relay link when a $4-$QAM constellation is used and the SNRs over the relay-destination and source-destination links are respectively given by $20$ dB and $-10$ dB. Comparisons are made with respect to the NAF algorithm in which no operation is performed at the source and relay and with the solution illustrated in \cite{Rong2009LinearRelayCommunication} and discussed in Section II in which the direct link is neglected. As seen, some improvement is achieved taking into account the direct link. 

\begin{figure}[tbp]
\centering \includegraphics[width=.44\textwidth]{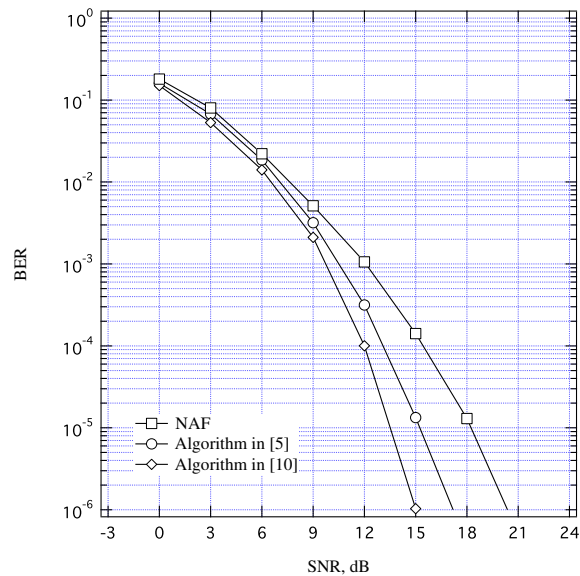}
\caption{BER of a one-way two-hop MIMO system as a function of the SNR over the source-relay link in the presence of direct link when $K= N_S =2$ and $N_R = 6$. In addition, the SNRs over the relay-destination and source-destination links are fixed to $20$ dB and $-10$ dB, respectively.}\label{fig10}
\end{figure}

Non-linear architectures in the presence of the direct link have recently been investigated in \cite{Tseng2011_TVTDFE} and \cite{Tseng2011_TVTTHP}. In particular, in \cite{Tseng2011_TVTDFE} a DFE-based system is considered. As for the linear case, it turns out that the joint optimization of the processing matrices is hard to address. To overcome this difficulty, the authors propose a suboptimal solution in which the precoding matrix $\bf{U}$ is constrained to have a unitary structure. On the other hand, in \cite{Tseng2011_TVTTHP} THP is employed at the source and only two suboptimal approaches are proposed. The first operates in an iterative manner and relies on the THP design proposed in \cite{Morelli2008} and on the relay precoder scheme derived in \cite{Tseng2010}. The second one provides a closed-form solution for the structure of ${\bf{U}}$ and ${\bf{F}}$ but it is again based on the assumption that ${\bf{U}}$ is unitary.

\section{Optimization of a two-way two-hop MIMO system}\label{Two-wayTwo-hop}

As seen, a one-way two-hop system requires four-phases to exchange the information between source and destination.
Since orthogonal channels (in time- or frequency-domain) are used to implement each phase, such a system spends twice as much channel resources with respect to a direct communication. To reduce this penalty, the two-way protocol originally proposed in \cite{Wittneben2007} (see also \cite{Tarokh2008} and references therein) allows source (user 1) and destination (user 2) to simultaneously transmit during the first phase. The received information at the relays is then forwarded to user 1 and 2 in the second phase. Since both user 1 and user 2 know their own transmitted data, they can remove the self-interference from the received signal provided that the required channel state information is available. This leads to a scheme with the same spectral efficiency of direct communications and at the same time able to take advantage of the potential benefits of relay communications.

\begin{figure}[tbp]
\centering \includegraphics[width=.45\textwidth]{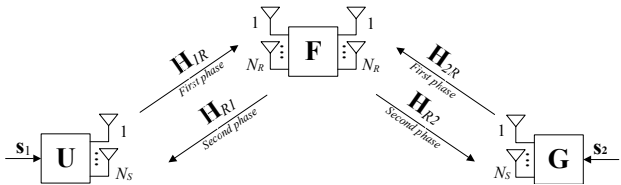}
\caption{Block diagram of two-way two-hop linear MIMO systems.}\label{fig4}
\end{figure}

The block diagram of a two-way two-hop MIMO relay network is shown in Fig. \ref{fig4} in which $\mathbf{s}_i$ for $i=1,2$ is used to denote the symbol vector transmitted by the $i$th user. In these circumstances, the signal received at the relay during the first phase can be expressed as
\begin{equation}\nonumber
\mathbf{y}_R = \mathbf{H}_{1R}\mathbf{U}\mathbf{s}_1 + \mathbf{H}_{2R}\mathbf{G}\mathbf{s}_2 +  \mathbf{n}_R
\end{equation}
where $\mathbf{H}_{iR}\in \CC^{N_R \times N_S}$ for $i=1,2$ denotes the channel matrix from the $i$th user to the relay while $\mathbf{n}_R$ accounts for thermal noise with zero mean and covariance matrix given by $\rho \mathbf{I}_{N_R}$. At the relay, the vector $\mathbf{y}_R$ is first processed by a matrix $\mathbf{F} \in \CC^{N_R \times N_R}$ and then forwarded to both users during the second phase. The received vector $\mathbf{y}_i$ for $i=1,2$ takes the form
\begin{equation}\label{10.2}
\mathbf{y}_1 =   \underbrace{\mathbf{H}_{R1} \mathbf{F}\mathbf{H}_{2R}\mathbf{G}\mathbf{s}_2}_{\text{useful signal}} +  \underbrace{\mathbf{H}_{R1} \mathbf{F}\mathbf{H}_{1R}\mathbf{U}\mathbf{s}_1}_{\text{self-interference}} + \underbrace{\mathbf{H}_{R1} \mathbf{F}\mathbf{n}_R + \mathbf{n}_1}_{\text{noise contribution}}
\end{equation}
and
\begin{equation}\label{10.3}
\mathbf{y}_2 =  \underbrace{\mathbf{H}_{R2} \mathbf{F}\mathbf{H}_{1R}\mathbf{U}\mathbf{s}_1}_{\text{useful signal}} +  \underbrace{\mathbf{H}_{R2} \mathbf{F}\mathbf{H}_{2R}\mathbf{G}\mathbf{s}_2}_{\text{self-interference}} + \underbrace{\mathbf{H}_{R2} \mathbf{F}\mathbf{n}_R + \mathbf{n}_2}_{\text{noise contribution}}
\end{equation}
where $\mathbf{H}_{Ri} \in \CC^{N_S \times N_R}$ for $i=1,2$ denotes the channel matrix from the relay to the $i$th user and $\mathbf{n}_i$ for $i=1,2$ is Gaussian with zero mean and covariance matrix $\rho \mathbf{I}_{N_S}$. As explained earlier, from \eqref{10.2} it is seen that if user 1 has knowledge of $\mathbf{H}_{R1} \mathbf{F}\mathbf{H}_{1R}$ it may remove its self-{interference} term from the received signal. The same can be done by user 2 if $\mathbf{H}_{R2} \mathbf{F}\mathbf{H}_{2R}$ is known. Then, the vectors at the input of the decision devices are given by $\mathbf{z}_1 = \mathbf{H}_{1}'\mathbf{G}\mathbf{s}_2 + \mathbf{n}_1' $ and $\mathbf{z}_2 = \mathbf{H}_{2}'\mathbf{U}\mathbf{s}_1 +  \mathbf{n}_2'$
where we have defined $\mathbf{H}_{1}' = \mathbf{H}_{R1} \mathbf{F}\mathbf{H}_{2R}$ and $\mathbf{H}_{2}' = \mathbf{H}_{R2} \mathbf{F}\mathbf{H}_{1R}$ while $\mathbf{n}_1'$ and $\mathbf{n}_2'$ are respectively given by $\mathbf{n}_1' = \mathbf{H}_{R1} \mathbf{F}\mathbf{n}_R + \mathbf{n}_1$ and $\mathbf{n}_2' = \mathbf{H}_{R2} \mathbf{F}\mathbf{n}_R + \mathbf{n}_2$.

Some recent works on two-way two-hop MIMO relay networks can be found in
\cite{Lee2010} and \cite{Hua2011} (see also \cite{Shin2010} and references therein). In particular, in \cite{Lee2010} the authors concentrate either on the maximization of the achievable sum rate or on the minimization of the sum of the MSEs while imposing power constraints at the terminals and relay. In both cases, they develop an iterative algorithm based on the gradient descendent technique that allows to numerically compute an approximation of the optimal $(\bf{G},\bf{U},\bf{F})$ (even for the case in which multiple relays are used). On the other hand, in \cite{Hua2011} the authors deal only with the maximization of the achievable sum rate and demonstrate that when $N_R \ge 2 N_S$ the optimal $\bf{F}$ takes the form
\begin{equation}\label{10.1}
\mathbf{F} = \mathbf{Q}_1\mathbf{A}\mathbf{Q}_2^H
\end{equation}
where $\mathbf{A} \in \CC^{2N_S \times 2N_S}$ is an arbitrary matrix while $(\mathbf{Q}_1,\mathbf{Q}_2)\in \CC^{N_R \times 2N_S}$ are semi-unitary matrices obtained from the following two QR decompositions:
\begin{equation}\nonumber
\left[\mathbf{H}_{R1}^H\;\mathbf{H}_{R2}^H\right] = \mathbf{Q}_1\mathbf{R}_1 
\end{equation}
and
\begin{equation}\nonumber
\left[\mathbf{H}_{1R} \; \mathbf{H}_{2R}\right]= \mathbf{Q}_2\mathbf{R}_2
\end{equation}
with $(\mathbf{R}_1,\mathbf{R}_2)\in \CC^{2N_S \times 2N_S}$ being upper triangular. On the basis of the above result, it follows that the original problem reduces to jointly design $(\bf{G},\bf{U},\bf{A})$. Clearly, this does not change the nature of the problem but it has the only practical relevance to lead to a simplification of the problem in those applications for which $N_R > 2 N_S$. Unfortunately, in \cite{Hua2011} it is shown that the optimal structure of the processing matrices $(\bf{G},\bf{U},\bf{A})$ is hard to find in closed-form since the optimization problem is not convex. This is in sharp contrast to the one-way relay systems discussed in Section II in which the optimal structure of $(\bf{G},\bf{U},\bf{F})$ are found in closed-form and leads to the diagonalization of the entire relay system. To overcome this difficulty, a couple of numerical methods based on iterative procedures in which $\bf{A}$ and $(\bf{G},\bf{U})$ are alternatively optimized are proposed in \cite{Hua2011}. In particular, the optimization of $\bf{A}$ for a given set of matrices $(\bf{G},\bf{U})$ is performed by means of two different algorithms. The first one is a hybrid algorithm in which the gradient descendent search and the {Newton}'s method are adaptively combined while the second one is inspired by the weighted minimum MSE algorithm originally proposed in \cite{Christensen2008}. On the other hand, the optimization of $(\bf{G},\bf{U})$ for a given $\bf{A}$ requires to solve a convex problem whose solution is found by means of the generalized waterfilling algorithm developed in \cite{Hua2010}.

From the above discussion, it follows that channel knowledge in two-way relay systems plays a key role since it is not only necessary for the design of the processing matrices but also for self-cancellation purposes. 
Although the problem of channel estimation in one-way relay systems has gained some interest, little work has been done for the two-way relay protocol. The problem has been recently investigated in \cite{Gao2009} in which a two-phase training-based algorithm is derived according to two different criteria. The first one relies on the maximum-likelihood methodology while the second one is derived so as to maximize the SNR at the receiver after taking the channel estimation errors into account. A possible drawback of these schemes is that they do not perform channel estimation at the relay in which only a scaling operation is performed. This means that the required channel state information can be provided at the relay only via a feedback channel. A different approach is illustrated in \cite{Jiang2010} in which channel estimation is performed at the relay during the first phase and it is then used in the second one to properly allocate the power so as to improve the channel estimate quality at the
terminal nodes. An alternative approach using the parallel factor analysis is proposed in \cite{RongChannel}. 

\section{Extensions and future lines of research}
This tutorial has discussed the optimization of AF MIMO relay systems. A number of key architectures has been reviewed and investigated under different design criteria and operating conditions. As seen, the optimization of a one-way two-hop MIMO system is a well-understood issue when perfect knowledge of the channel matrices is assumed. Much work has still to be done for the development of robust solutions able to cope with channel uncertanties. On the other hand, different aspects in the optimization of other architectures remain unsolved and/or to be further investigated. An interesting problem common to all the investigated systems is the deployment of decentralized algorithms that may provide a good tradeoff between performance and system scalability.  
In addition to all this, there are still many extensions and future lines of research related to the optimization of MIMO relay networks, some of which are briefly described in the next. {For example, the optimization of multiuser MIMO
relay systems has only recently become an active research topic. In particular, the multiaccess MIMO
relay networks are addressed in \cite{RongMURelay} -- \cite{RongNoCSI} while the multiuser broadcasting relay systems are
investigated in \cite{Hua2010}. Compared to its single-user
counterpart, the optimization of a
multiuser system is much more challenging due to the considerable amount of required channel state
information and computational complexity. Another interesting direction
is the interference MIMO relay network where multiple source nodes
communicate with their desired destination nodes with the aid of
(distributed) relay nodes \cite{Tellambura2009} and \cite{Khandaker}. Due to the existence
of interference, a cross layer design between the MAC layer and
the PHY layer may be employed to optimize the overall system
performance. Also, the optimization of full-duplex relay networks in which loop-back interference is traded for higher spectral efficiency is worthy further
investigation \cite{HuaMilcom}.}

\end{document}